\documentclass[12pt, final]{article}
\PassOptionsToPackage{bookmarksopen,bookmarksdepth=2}{hyperref}
\usepackage{jheppub}
\usepackage[utf8x]{inputenc}
\usepackage{lmodern, microtype}
\usepackage{graphicx}
\usepackage{amsmath, amsfonts,amssymb,mathtools,dsfont,bm,nicefrac}
\usepackage{latexsym}
\usepackage[dvipsnames]{xcolor}
\usepackage{cancel,slashed,booktabs,soul}
\usepackage{nccmath, textcomp,bigints} 
\usepackage{stackengine,subcaption}
\usepackage{etoolbox}   % for booleans and much more
\usepackage{verbatim}   % for the comment environment
\usepackage[obeyFinal,backgroundcolor=white,linecolor=black,textsize=footnotesize]{todonotes}
\newcommand{\vev}[1]{\langle #1 \rangle}
\newcommand{\floor}[1]{\lfloor #1 \rfloor}
\DeclareMathOperator{\re}{Re}
\DeclareMathOperator{\im}{Im}
\DeclareMathOperator{\sgn}{sgn}
\DeclareMathOperator{\poly}{Li}
\newcommand{\Li}[2]{\poly_{#1}(#2)}
\DeclareMathOperator{\dd}{d \!}
\renewcommand{\d}{\dd \, \!}

\newcommand{\mcb}{m_{\rm \textsc{cb}}}

\newcommand{\oneloop}{\text{1-loop}}
\newcommand{\cs}{{\rm \textsc{cs}}}
\newcommand{\f}{a}
\newcommand{\gs}{{\rm \textsc{gs}}}
\newcommand{\af}{{\rm \textsc{af}}}
\newcommand{\pv}{{\rm \textsc{pv}}}
\newcommand{\vz}{\vev{\zeta}}

\newcommand{\Z}{\mathbb Z}
\newcommand{\qz}{q \, \vz}

\DeclareMathOperator{\sign}{sign}
\newcommand{\ffpi}{\frac{1}{4\pi}}

\newcommand{\fr}[1]{\underline{#1}}

\newcommand{\m}{\mu}

\renewcommand{\bm}{{\bar \mu}}
\newcommand{\bn}{{\bar \nu}}
\newcommand{\ba}{{\bar a}}
\newcommand{\bb}{{\bar b}}
\newcommand{\bt}{{\bar 3}}
\renewcommand{\c}{\mathcal }
\newcommand{\C}{C_3}
\renewcommand{\S}{\Sigma}

\renewcommand{\a}{\alpha}
\newcommand{\fm}{\floor{\m}}

\newcommand{\g}{\gamma}
\newcommand{\tr}[1]{{\rm Tr} \left( #1\right)}
\newcommand{\intdk}{\int \frac{\d^3k}{(2\pi)^3}}

\newcommand{\intk}{\int \frac{\d^3k}{2 \pi^2 }}
\newcommand{\sumintk}{\sum_{n\in \Z } \intk \sum_{s=0}^3}

\newcommand{\abs}[1]{\left | #1 \right |}
\newcommand{\sums}{\sum_{s=0}^3 (-1)^s \,}
\newcommand{\limms}{\lim_{M_s\to \infty}}
\newcommand{\limm}{\lim_{M\to \infty}}

\newcommand{\sumn}{\sum_{n \in \Z}}
\newcommand{\e}{\epsilon}

\newcommand{\reg}{{\rm reg}}
\newcommand{\regb}{{\rm \textbf{reg}}}

\newcommand{\bbn}{\boldsymbol{n}}
\newcommand{\mmfrac}[2]{\mfrac{\raisebox{-2pt}{$#1$}}{#2}} 
\newcommand{\mtfrac}[2]{\mfrac{\raisebox{-2pt}{$#1$}}{\raisebox{2pt}{$#2$}}}
\newcommand{\nfrac}[2]{\frac{\raisebox{-2pt}{$#1$}}{#2}} 

\title{\centering \large Chiral anomalies on a circle and their cancellation in  F-theory }

\author[]{Pierre Corvilain$^{a}$,}
\author[]{Thomas W.~Grimm$^{a}$}
\author[]{and Diego Regalado$^b$}

\affiliation[]{$^a$Institute for Theoretical Physics and \\
Center for Extreme Matter and Emergent Phenomena,\\
Utrecht University, Leuvenlaan 4, 3584 CE Utrecht, The Netherlands}
\affiliation[]{$^b$Theoretical Physics Department, CERN, Geneva, Switzerland}

\emailAdd{p.l.j.corvilain@uu.nl}
\emailAdd{t.w.grimm@uu.nl}
\emailAdd{diego.regalado@cern.ch}

\abstract{We study in detail how four-dimensional local anomalies manifest themselves when the theory is compactified on a circle. By integrating out the Kaluza-Klein modes in a way that preserves the four-dimensional symmetries in the UV, we show that the three-dimensional theory contains field-dependent Chern-Simons terms that appear at one-loop. These vanish if and only if the four-dimensional anomaly is canceled, so the anomaly is not lost upon compactification.
We extend this analysis to situations where anomalies are canceled through a  Green-Schwarz mechanism. 
We then use these results to show automatic cancellation of local anomalies in F-theory compactifications that can be obtained as a limit of M-theory on a smooth Calabi-Yau fourfold with background~flux.}

\renewcommand{\baselinestretch}{1.2} 

\definecolor{blue}{rgb}{0.06, 0.3, 0.57}

\apptocmd\normalsize{%
 \abovedisplayskip=12pt plus 3pt minus 6pt
 \abovedisplayshortskip=0pt plus 3pt 
 \belowdisplayskip=12pt plus 3pt minus 6pt
 \belowdisplayshortskip=7pt plus 3pt minus 4pt
}{}{}

\begin{document}

\makeatletter
\let\old@fpheader\@fpheader
\renewcommand{\@fpheader}{\old@fpheader\hfill
CERN-TH-2017-218}
\makeatother

\setlength{\parskip}{5pt}

\maketitle

%%%%%%%%%%%%%%%%%%%%%%%%%%%%%%%%%%%%%%%%%%%%%%%%
\section{Introduction} \label{sec:introduction}
%%%%%%%%%%%%%%%%%%%%%%%%%%%%%%%%%%%%%%%%%%%%%%%%

\renewcommand{\baselinestretch}{1.2} 

It is sometimes impossible to 
define the partition function of four-dimensional gauge theories due to the presence of local anomalies. These signal the breakdown of gauge invariance at one-loop and spoil unitarity. They can occur when chiral fermions charged under the gauge group are present, precisely because there is no UV regulator preserving both four-dimensional Lorentz and gauge invariance (for a pedagogical review, see~\cite{Bilal:2008qx}).

If we consider an anomalous theory, not on $\mathbb R^4$, but on $\mathbb R^3\times S^1$, the behavior at short distances is exactly the same, so one faces the same problems in trying to define the theory, which is equally inconsistent. Thus, one cannot expect the theory to be well-defined for energies below the (inverse) compactification radius, even though the effective theory is three-dimensional --- which does not allow for local gauge anomalies. In this paper we show explicitly that the effect of having a local gauge anomaly in four dimensions, appears as a field-dependent one-loop Chern-Simons term in three dimensions. In order to do so, we use a regulator for the four-dimensional theory on $\mathbb R^3\times S^1$ that preserves \emph{four-dimensional} Lorentz invariance at high energies. If we were to use a regulator that preserves only three-dimensional Lorentz invariance, we would not find these field-dependent Chern-Simons terms.
This, however, would not correspond to a theory which is four-dimensional at high energies.

\medskip

We first illustrate this in the simplest example, i.e.~a chiral fermion charged under an Abelian gauge field on a circle. To begin with, we present a detailed account of the circle Kaluza-Klein reduction from four to three dimensions. Starting with a possibly anomalous four-dimensional theory, we show that integrating out the Kaluza-Klein modes generates a three-dimensional Chern-Simons term at one-loop, which depends continuously on the Coulomb branch parameter when anomalies are not canceled. We see that the precise form of this term depends crucially on the regularization scheme and argue that the proper scheme to pick is the one that respects the four-dimensional space-time symmetries. We also discuss the case in which the four-dimensional theory cancels the anomaly via a Green-Schwarz mechanism.

We then extend the discussion to the case in which the fermions are coupled to a background metric. More precisely, we present a detailed computation of the Chern-Simons terms involving the Kaluza-Klein vector of the four-dimensional metric along the circle. Again we observe that one-loop Chern-Simons terms are gauge invariant if and only if four-dimensional local gauge anomalies are canceled. Even though we do not consider the mixed gauge-gravitational anomaly explicitly, we expect it to appear as a field-dependent gravitational Chern-Simons term in three dimensions, similarly to the cubic gauge anomaly.

As an application of this result, we show that local gauge anomalies cancel for four-dimensional F-theory compactifications that can be obtained as a limit of M-theory on a smooth Calabi-Yau fourfold with background flux. F-theory is often defined as a non-perturbative version of Type IIB string theory with seven-branes~\cite{Vafa:1996xn}. With our current understanding of this formulation, however, crucial questions especially concerning the effective actions arising in F-theory cannot be easily settled. Clearly, checking whether anomalies cancel is one of the most basic properties that can be analyzed for an F-theory effective action. Focusing only on weakly coupled Type IIB string theory with D7-branes and O7-planes, it is well understood that the D-branes and fluxes need to be chosen such that they cancel tadpoles, see e.g.~\cite{MarchesanoBuznego:2003axu,Blumenhagen:2006ci,Plauschinn:2008yd}. An analogous understanding for general seven-brane settings in F-theory is still lacking. Furthermore, at weak coupling, the demand for tadpole cancellation can be technically hard to implement, since it involves various choices for the D7-branes and flux. In contrast, the geometric approach to F-theory using elliptic fibrations already selects configurations that satisfy certain consistency conditions, which sometimes are not easy to interpret. For example, the anomaly conditions can be translated into geometric relations~\cite{Grassi:2011hq,Park:2011ji,Cvetic:2012xn,Bies:2017abs} that can be checked for individual geometries but have not been proved generally. It is thus a pressing question, how exactly anomaly cancellation in the F-theory effective actions is encoded in the choice of geometry and flux.

To approach four-dimensional F-theory settings we use the fact that, when compactified on a circle, it is dual to M-theory compactified on a Calabi-Yau fourfold with flux. Thus, when the fourfold can be smoothed out, we may compute the three-dimensional effective action by dimensional reduction of eleven-dimensional supergravity~\cite{Denef:2008wq,Grimm:2010ks}. Using the results in the previous sections, we can rule out anomalies of the four-dimensional theory by simply looking at the three-dimensional effective action obtained in this way, and checking whether it contains field-dependent Chern-Simons terms. We find that, as expected, anomalies are canceled for this large class of models. This extends and clarifies the arguments given in~\cite{Grimm:2015zea,Grimm:2015wda}, where F-theory anomaly cancellation was studied only using large gauge transformations around the circle.

We would like to point out that if one is only interested in studying how the four-dimensional anomalies arise after compactification, one may infer the three-dimensional (field-dependent)  Chern-Simons terms in the effective action by demanding they reproduce the anomaly~\cite{Jensen:2013kka,Jensen:2013rga,DiPietro:2014bca}. However, in order to compute the full effective action, it is necessary to use a regularization scheme like the one presented here. Thus, the fact that our computation yields such three-dimensional Chern-Simons terms can be regarded as a check of the technique itself which can then be used to compute more general quantities. In particular, it may be used in the context of F-theory compactifications to explore higher order corrections beyond anomalies. 

The paper is organized as follows. In section~\ref{sec:aom_circle} we consider compactification on a circle of a chiral fermion coupled to a $U(1)$ gauge field.
We discuss in detail the particular regulator that we use and compute the Chern-Simons terms that are generated at one-loop. We extend this analysis to a theory that includes a Green-Schwarz term. In section~\ref{sec:anom_circle_grav} we follow a similar strategy including the Kaluza-Klein vector, allowing us to partially treat the coupling to gravity. Our most general results are presented at the end of this section, where we discuss the case with multiple fermions coupled to a general gauge group, and multiple Green-Schwarz scalars. Finally, in section~\ref{sec:M-theory} we use these results to discuss the anomalies in four-dimensional effective theories coming from F-theory. Comparing the circle compactified theories with the dual M-theory description we show that local gauge anomalies are automatically canceled. We include many details of the computations in the appendices.

%2%%%%%%%%%%%%%%%%%%%%%%%%%%%%%%%%%%%%%%%%%%%%%%%
\section{4D Chiral anomaly on a circle} \label{sec:aom_circle}
%%%%%%%%%%%%%%%%%%%%%%%%%%%%%%%%%%%%%%%%%%%%%%%%

\subsection{Classical compactification}
\label{chiral_circle}

We start by performing the classical circle compactification of four-dimensional chiral fermions coupled to a gauge field. For simplicity we will consider for the moment just one left-handed fermion charged under a single $U(1)$ gauge group. The action is
\begin{align}\label{4dact}
S_4 = \int \bar {\hat \psi} \,\hat \gamma^\m \left( i   \partial_\m + q \hat g \hat{A}_\mu \right) P_L \,\hat \psi \star 1 \,,
\end{align}
where $\hat F = \d \hat A$ is the field strength of the gauge field $\hat A$, whose coupling constant is $\hat g$. $\hat \psi$ is a Dirac fermion, with charge $q$ under $\hat A$, and $P_L = \frac 12 (1+ \gamma^5)$ is the chirality projector.
Four-dimensional quantities are denoted by a hat; later on, unhatted objects will be three-dimensional. We work with the `mostly minus' signature.

As is well-known\footnote{For a pedagogical review, see~\cite{Bilal:2008qx}.}, the theory~\eqref{4dact} is anomalous, namely, the one-loop quantum effective action transforms under a gauge transformation, $\hat A \to \hat A + \d \hat \lambda$, as
\begin{align}\label{var}
\delta S_4^{\rm 1PI} = \frac{\hat g^3}{24 \pi^2} \,\, q^3 \int \hat \lambda \, \hat F \wedge \hat F\,.
\end{align}
Including multiple left-handed fermions, one finds that the theory is anomaly free when $\sum_\f q^3_\f =0$, with $\f$ running over the fermions. Since we are interested in showing the effects of having an anomaly when the theory is compactified on a circle, it is enough to consider a single fermion.

We take the space-time to be of the form $\mathbb R^3 \times S^1$ with metric
\begin{equation}\label{metric}
\dd\hat s^2=\dd s^2-r^2\dd y^2.
\end{equation}
Here $\d s^2$ is the flat metric in three dimensions and $y\sim y+2\pi$ is a coordinate on the circle of radius $r$.
The vector $\hat A$ yields in $\mathbb R^3$ a Kaluza-Klein (KK)  tower of vectors $A_n$ and scalars $\zeta_n$. The fermion $\hat \psi$ gives a KK tower of fermions $\psi_n$. 
The massive vector and scalar modes, $A_n$ and $\zeta_n$, will not play any role in our discussion, essentially because they are not chiral and the anomaly is induced by chirality. The fermionic KK modes, on the other hand, are essential. Thus, we take the reduction Ansätze of the four-dimensional fields to be
\begin{align}
\hat A (x,y) &= \mfrac{1}{\sqrt {2 \pi \,  r}} \, A(x) + \mmfrac 1{\hat g}\, \zeta(x) \dd y + \ldots \,, \label{ansatzA}\\
\hat \psi (x,y) &= \mfrac{1}{\sqrt {2\pi  \, r}}\sum_{n\in \mathbb Z} \psi_n (x) \, e^{-in y} \,,
\end{align}
where the dots stand for the massive KK modes of the vector $\hat A$ that we will ignore in the following.
With these Ansätze, the three-dimensional action that one obtains from~\eqref{4dact} is 
\begin{align}\label{3dact}
S_3 = \sum_{n\in\mathbb Z} \int \bar \psi_n  \left  [\gamma^a \!\left(i\partial_a + q g  A_a  \right)+  \mtfrac 1r \gamma^3 \!\left (n + q  \zeta\right ) \right ] \!P_L \, \psi_n \star 1,
\end{align}
where $g$ is the three-dimensional effective coupling constant, which is defined as
\begin{equation}
g  = \frac{\hat g}{\sqrt {2\pi \, r}}\,.
\end{equation}
Notice that the fermions in~\eqref{3dact} are still 4-components spinors (but the projector $P_L$ effectively removes 2 components), and the gamma matrices are still $4 \times 4$ matrices, defined by $\hat \gamma^\m = (\g^a, \frac 1r \g^3)$. We could switch to a formulation with 2-components spinors and   $2\times 2$ gamma matrices~\cite{Appelquist:1986fd}. However, it will be more convenient for us to keep working with this 4-components formulation.

From the action~\eqref{3dact}, one reads off the mass of $\psi_n$ to be $n/r$, as expected.
In addition to this, in the Coulomb branch, i.e.~when we give the scalar a vev $\zeta = \vz + \chi$, each mode receives an extra contribution to its mass, proportional to the vev, namely
\begin{equation}
m_{\rm \textsc{cb}} = \mtfrac 1r \, q \, \vz \,.
\end{equation}
This means that for generic values of $\langle \zeta\rangle$ even the zero mode becomes massive. From now on, we will assume that $\langle \zeta \rangle$ is non-zero and we will be interested in the Wilsonian effective action at an energy scale lower than $\mcb$ so the whole tower of fermions has to be integrated out.

Let us now briefly discuss the symmetries of the three-dimensional theory. Since we have a $U(1)$ gauge field on $\mathbb R^3 \times S^1$, the four-dimensional theory admits both small and large gauge transformations. More explicitly, $\d \hat \lambda (x,y)$ decomposes as\footnote{Here we also ignore the higher modes of $\hat\lambda$ which correspond to the gauge parameters of the massive gauge bosons.}
\begin{equation}
\d \hat \lambda (x,y) = \mfrac{1}{\sqrt{2\pi r} } \d \lambda (x) + \mfrac{1}{\hat g} \, n \d y\,.
\end{equation}
From the three-dimensional perspective, the first term corresponds to gauge transformations of the massless vector, $A \to A + \d \lambda$, while the second term translates into a discrete shift symmetries of the scalar,
\begin{equation}\label{shiftz}
\zeta \to \zeta +n\,.
\end{equation}
As can be seen from~\eqref{var}, the four-dimensional theory is not invariant under both these transformations, and therefore the three-dimensional theory is also expected to be \emph{non}-invariant under both their three-dimensional counterparts. This will be a guiding principle for the remainder of this paper. We may compute the expected variation of the three-dimensional one-loop effective action by formally reducing~\eqref{var} on the circle,
\begin{equation}\label{var3d}
\delta S_3^{\, \rm eff} =  \frac{g^2}{12 \pi} \, \, q^3 \int n \, A \wedge F - 2 \,  \chi  \dd \lambda \wedge F \,.
\end{equation}
That is to say, we expect the three-dimensional effective action obtained after integrating out all the massive fermions to vary as~\eqref{var3d} under the variations $\delta A = \d \lambda$ and $ \delta \zeta = n$.
We now turn to computing this effective action.

% % % % % % % % % % % % % % % % % % % % % % % % % % % % % % % % % % % % % % % %
\subsection{Integrating out the Kaluza-Klein modes}
\label{sec:3dreg}

As explained in the previous subsection, we would like to compute the Wilsonian effective action at a scale below $\mcb$, obtained by integrating out the whole KK tower of fermions. It is well-known that integrating out a single fermion of mass $m$ and charge $q$ in three dimensions generates a CS coupling, $ \frac{g^2}{4\pi} \Theta \, A\wedge F$, with the CS coefficient being\footnote{The loop computation giving this result is included in appendix~\ref{ap:AA}.}
\begin{equation}\label{int_ferm}
\Theta = \mmfrac 12   \, q^2 \sign(m) \,.
\end{equation}
Since the $n$-th KK mode has a mass $m_n = \frac 1r (n + \qz)$, integrating out the whole tower gives 
\begin{equation} \label{sum}
\Theta = \mmfrac 12  \, q^2 \sum_{n \in \Z} \sign \left (n + \qz \right ) \,.
\end{equation}
This sum is divergent and can be regularized, for instance, using the zeta function regularization, as is done in~\cite{Poppitz:2008hr,Golkar:2012kb,DiPietro:2014bca,DiPietro:2016ond} (see appendix~\ref{ap:Sums} for a detailed discussion of the regularization).\footnote{The use of zeta function regularization in the study of anomalies and dimensional reduction has already been investigated in the past, see for instance~\cite{Duff:1982gj,Duff:1982wm}.} The result of the regularization is\footnote{One can also obtain this result by introducing a set of three-dimensional Pauli-Villars  particles for each KK mode, which render the sum~\eqref{sum} finite. This is done in appendix~\ref{ap:AA}.}
\begin{equation}\label{summed}
\Theta^{\reg} = q^2 \left( \mmfrac 12   + \floor{\qz } - \qz \right)\,.
\end{equation}
While each term of the sum~\eqref{sum} is quantized,
the result of the sum is not since the last term of~\eqref{summed} depends linearly on the continuous parameter $\vz$. 
This is due to the regularization and is a manifestation that our original theory was four-dimensional. Notice that the continuous term is proportional to $q^3$, and would vanish for an anomaly free theory.
This non-quantized CS coefficient is therefore a three-dimensional manifestation of the four-dimensional anomaly. 

Notice that the result~\eqref{summed} is invariant under the shift~\eqref{shiftz}. However, as we saw in the previous section, the three-dimensional effective action is not expected to be invariant under such shifts, cf.~\eqref{var3d}. More generally, as already explained in the introduction, we would like to regulate the theory in a way that respects the symmetries of the four-dimensional theory at high enough energies. It is not clear a priori whether the zeta function regularization satisfies this criterion or not. In the following, we will regulate the theory in a different way, which makes it clear that the result is compatible with the four-dimensional symmetries in the UV. The result will not be invariant under discrete shifts of the Coulomb branch parameter~\eqref{shiftz}, in agreement with~\eqref{var3d}. 

% % % % % % % % % % % % % % % % % % % % % % % % % % % % % % % % % % % % % %
\subsection{Integrating out the Kaluza-Klein modes, preserving the 4D symmetries}

In order to regulate the three-dimensional effective action while preserving four-dimen\-sional Lorentz invariance, the natural thing to do is to reduce a four-dimensional regulator on a circle, so we briefly review a four-dimensional regularization procedure.

The regulator we consider in four dimensions is a non-gauge invariant version of the Pauli-Villars (PV) regularization. It normally consists in introducing a set of new particles with possibly different statistics and large mass $M_s$. 
However in our case~\eqref{4dact} chirality prevents from adding a mass term, so one has to include a right-handed fermion as well, which does not couple to the gauge field, therefore breaking gauge invariance~\cite{AlvarezGaume:1984dr} (see~\cite{Bilal:2008qx} for a pedagogical review). This is precisely the anomaly: one is able to regulate the theory only if one gives up on gauge invariance. 
Concretely, we replace the original Lagrangian~\eqref{4dact} by 
\begin{equation}\label{S4PVs}
S_4^\pv =\int \sum_{s=0}^3 (-1)^s \,  \bar{\hat{\psi}}_s \left[ \hat \gamma^\m \left( i { \partial}_\mu   + q \, \hat g  \,\hat{ A }_\m \,  P_L \right)  + M_s \right] \hat \psi_s \, \, \hat \star \, 1 \,,
\end{equation}
with $M_0=0$ and $\sum_{s=1}^3 (-1)^s M^2_s =0$, and then take the limit in which $M_s$ goes to infinity for $s=1,2,3$. The case $s=0$ formally defines the same theory as~\eqref{4dact}.\footnote{That is, it generates the same perturbative expansion.} Notice that this Lagrangian breaks gauge invariance explicitly which (for an anomalous theory as the one we are analyzing) persists even in the limit where we decouple the PV particles. In particular, the triangle diagram using this Lagrangian gives the consistent anomaly~\cite{AlvarezGaume:1984dr}.

Reducing the regularized theory~\eqref{S4PVs} on a circle using the Ansätze~\eqref{ansatzA}, we find
\begin{equation}\label{3dactPV}
S_3^\pv = \sum_{n \in \Z}  \int \sum_{s=0}^3 (-1)^s  \, \bar{ \psi}_{n,s} \left[ \g^a \left(i { \partial }_a   + q g { A }_a P_L\right) 
  + \mtfrac 1r \gamma^3( n + q\, \zeta  P_L)
  + M_s\right] \psi_{n,s}  \star 1 \,,
\end{equation}
For $s=0$, this theory is formally the same as the one given in~\eqref{3dact}. The three other KK-towers, for which $M_s\neq 0$, provide a three-dimensional regulator that preserves four-dimensional Lorentz invariance (by construction), but explicitly breaks small gauge transformations as well as the shift symmetry of the scalar. 

\begin{figure}
\centering
\begin{subfigure}{0.4\textwidth}
\centering
\includegraphics[scale=1.15]{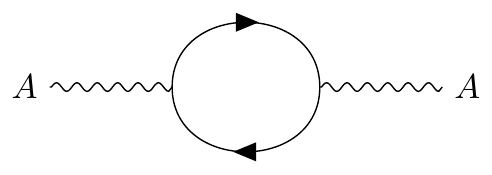}
\caption{}
\label{fig:a}
\end{subfigure}
\hspace{3mm}
\begin{subfigure}{0.4\textwidth}
\centering
\includegraphics[scale=1]{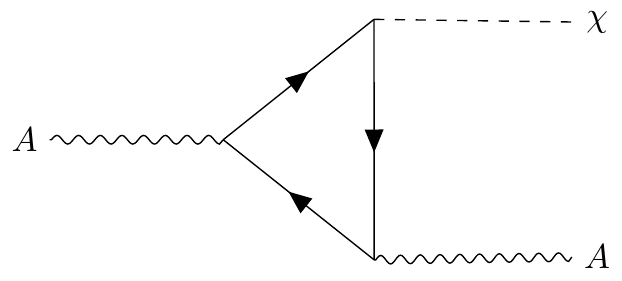}
\caption{}
\label{fig:b}
\end{subfigure}
\caption{One-loop diagrams contributing to the field independent (a) and field dependent (b) Chern- Simons terms in three dimensions.}
\label{fig:oneloopCS}
\end{figure}

At this point, we may compute the one-loop diagrams of figure~\ref{fig:oneloopCS} using the Feynman rules derived from~\eqref{3dactPV} to find the effective action in three dimensions. The details of this computation are presented in appendix~\ref{ap:AA}, where we find that
\begin{equation}\label{3dcons}
S_3^{\rm eff}= \frac{\raisebox{-2pt}{$1$}}{4\pi}g^2\int \Theta^{\regb}\, A\wedge F\,,
\end{equation}
with
\begin{equation} \label{PVreg}
\Theta^{\regb}= q^2 \left( \mmfrac 12   + \floor{q\,\zeta } - \mmfrac 23 \, q\,\zeta \right) \,.
\end{equation}
The bold superscript $\regb$ indicates that we used a different regularization than before. Notice that we included both the constant term coming from the diagram~\ref{fig:a} as well as the field-dependent term coming from the triangle diagram~\ref{fig:b}. Thus, we find a field-dependent CS coefficient which violates gauge invariance. In particular, under a (small) gauge transformation of the three-dimensional gauge field, $A\rightarrow A+\d\lambda$, as well as a discrete shift of the scalar, $\zeta\rightarrow \zeta+n$, we find precisely~\eqref{var3d}.

% % % % % % % % % % % % % % % % % % % % % % % % % % % % % % % % % % % % %
\subsection{Anomaly Inflow}
\label{sec:AI}
We have presented two different ways of regularizaring the loop diagrams, the first one yielding a gauge invariant effective action, and the second one a non gauge-invariant effective action. We now briefly discuss how the two are related, following~\cite{Jensen:2013kka,DiPietro:2014bca}.

The chiral theory we started with in subsection~\eqref{chiral_circle} is non gauge-invariant at one-loop, with variation given by~\eqref{var}. This variation can be canceled by adding a term in the action of the form
\begin{equation}\label{CS5}
S_5=-\frac{\hat g^3}{24 \pi^2}\,\, q^3 \int_{\mathcal{M}_5}  A^{(5)} \wedge F^{(5)} \wedge F^{(5)} \,,
\end{equation}
where $ \partial \mathcal{M}_5 =  \mathbb{R}^4 $ and the field $A^{(5)}$ is an extension of the four-dimensional gauge field to $\mathcal M_5$. Under a gauge transformation we have that
\begin{equation}\label{delCS5}
\delta S_5= -\frac{\hat g^3}{24 \pi^2}\,\, q^3 \int_{\mathbb R^4} \hat \lambda \, \hat F\wedge \hat F \,,
\end{equation}
which cancels~\eqref{var}. This means that if one considers the four-dimensional theory~\eqref{4dact} together with this 5D CS term~\eqref{CS5}, the full theory is gauge invariant. For constant~$\zeta\,(=\vz)$, we may compactify this gauge invariant action on a circle, which leads to the covariant effective action, as opposed to the consistent one given in~\eqref{3dcons}. Taking $\c M _5 = \c N_4 \times S^1$, with $\partial \c N_4 = \mathbb R^3$, we find that
\begin{equation} \label{AI}
S_5= -\frac{g^2}{12 \pi} \, q^3 \, \int_{\mathbb R^3 } \vz \, A \wedge F\,,
\end{equation}
which is precisely the difference between the result obtained by using the zeta-function regularization~\eqref{summed} and the (non gauge invariant) PV regularization~\eqref{PVreg}, for constant~$\zeta$. Thus, the former one gives the covariant effective action, while the latter gives the consistent effective action, the two being related through anomaly inflow, 
\begin{equation}\label{thetasrel}
\Theta^\reg = \Theta^\regb + \Theta^{\rm infl} \,.
\end{equation}

% % % % % % % % % % % % % % % % % % % % % % % % % % % % % % % % % % % % % %
\subsection{Chern-Simons coefficient for an anomaly-free theory}
\label{sec:CS}

Let us consider now the case in which anomalies are canceled, i.e. $\sum_\f q_\f^3=0$. In that case, the consistent and covariant effective actions coincide (the anomaly inflow piece vanishes), and are gauge invariant, as expected. The CS coefficient is then
\begin{equation}
\Theta^\af = \sum_{a =1}^{N_f}  q_a^2 \left( \floor{q_a \vz } + \mmfrac 12    \right ) \label{theta_af} \,,
\end{equation}
where we included $N_f$ fermions $\psi_\f, \ \f=1,\ldots,N_f$, with charges $q_\f$.
At first sight this CS coefficient might not look integer since it includes the sum of half-integer numbers. However, the anomaly conditions ensures that this does not happen. Indeed, since $q_a^3\equiv q_a^2 \mod 2$ for all $a$, we have that $\sum_a q_a^3 =0$ implies $\sum_a q_a^2 \equiv 0 \mod 2$, so the CS coefficient is integer. Furthermore, the anomaly condition makes the result invariant under discrete shifts of the Coulomb branch parameter. Indeed, under a shift $\zeta \to \zeta + n$, the variation of~\eqref{theta_af} is $\sum_a q_a^3$ and therefore vanishes.\footnote{In fact, the period of $\Theta^\af$ is in general $1/L$, where $L={\rm gcd}(q_\f)$, which reflects a global $\mathbb Z_L$ discrete symmetry.}

\begin{figure}[h!]
\centering
\includegraphics[width=0.85\linewidth]{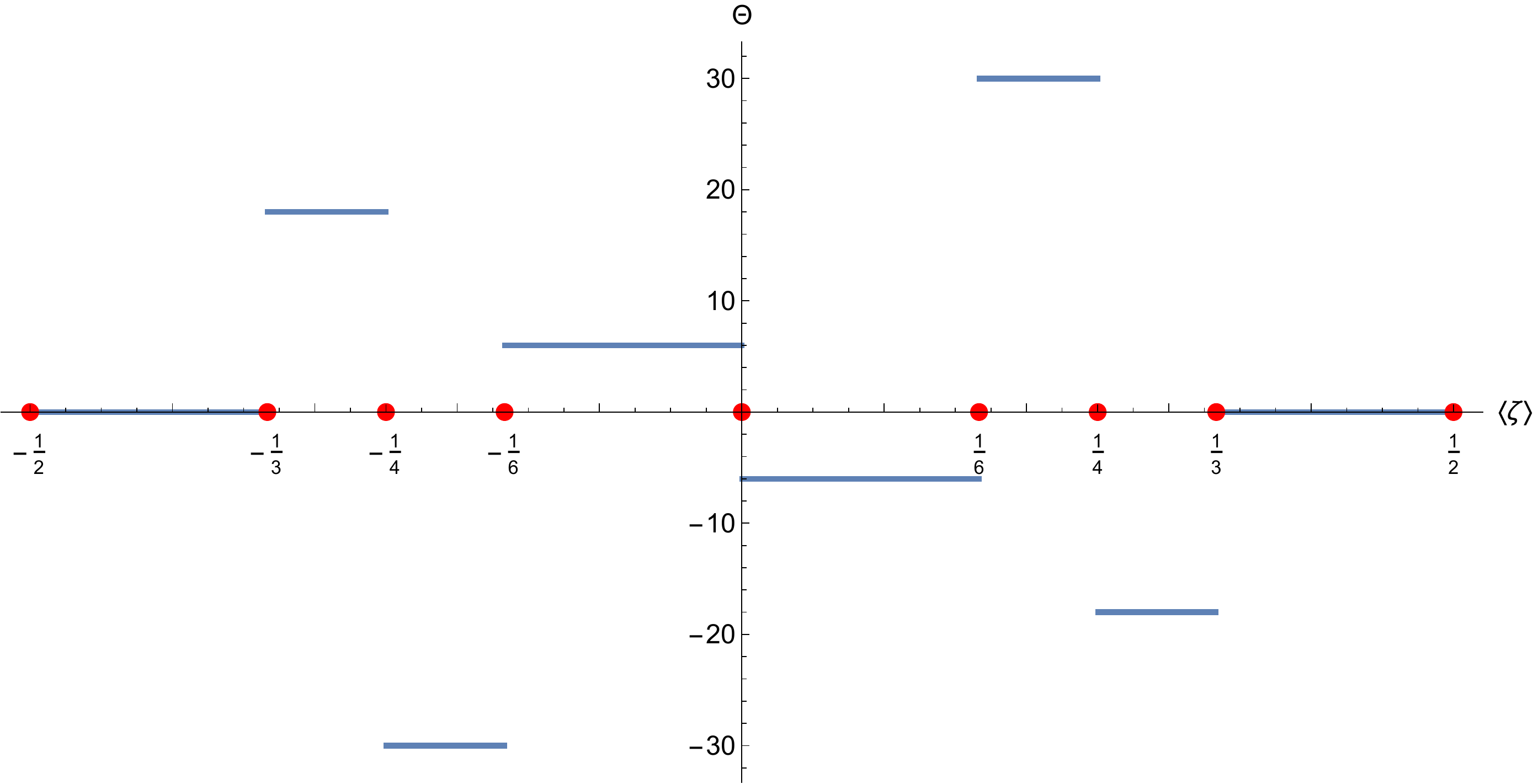}
\caption{Plot of the value of the CS coefficient $\Theta^\af$, given in~\eqref{theta_af}, with charges being the hypercharges of the Standard Model, as a function of the Coulomb branch parameter $\vz$, over the fundamental period $\vz \in \left [-\frac 12 , \frac 12 \right ]$. The red points denote the values of $\vz$ for which some $q_\f \, \vz \in \mathbb Z$, where at least a fermion is massless. 
}\label{example1}
\end{figure}

In figure~\ref{example1}, we plot the value of $\Theta^\af$ as a function of $\vz$ for a concrete example of a theory canceling the pure gauge anomaly, namely the fermions of the Standard Model charged under the hypercharge $U(1)$. The value of $\Theta^\af $ jumps at the points of the Coulomb branch where $q_\f \, \vz \in \mathbb Z$, at which at least one of the fermions becomes massless.
At those points, the Wilsonian effective action breaks down since we integrated out a massless field.

% % % % % % % % % % % % % % % % % % % % % % % % % % % % % % % % % % % % % % % %
\subsection{Green-Schwarz mechanism}
\label{sec:GS}

In this section we discuss the more general case in which the four-dimensional theory contains Green-Schwarz terms. These will be important when we discuss F-theory compactifications in section~\ref{sec:M-theory} since in that case one typically finds that anomalies are canceled by the appearance of such terms.

We denote by $\hat \rho$ the scalar field of the Green-Schwarz mechanism, which is gauged through the covariant derivative $\hat D \hat \rho = \dd\hat\rho - \theta \hat A$, from which one reads its variation under a gauge transformation $\delta \hat \rho = \theta \hat \lambda$. We then add to the action~\ref{4dact} the non gauge invariant term 
\begin{equation}\label{GSact}
S_4^\gs= \frac{\hat g^3}{24 \pi^2} \, \,C \int \hat\rho \, \hat F\wedge\hat F \,.
\end{equation}
In our conventions the scalar is periodic $\hat\rho\sim\hat\rho+1/\hat g$ and $\theta$ is integer. The variation of~\eqref{GSact} under a gauge transformation is 
\begin{equation}
\delta S_4^\gs = \frac{\hat g^3}{24 \pi^2} \,\, \theta \, C\int \hat\lambda \, \hat F\wedge\hat F\,,
\end{equation}
such that the anomaly cancellation condition (for multiple fermions) reads
\begin{equation} \label{AC}
\sum_a q_a^3 + \theta \,C= 0 \,.
\end{equation} 

We now reduce the  term~\eqref{GSact} on a circle using the following Ansatz for $\hat \rho$,
\begin{equation} \label{ansatz_rho}
\dd\hat\rho=\mfrac{1}{\sqrt{2\pi r}}\, \dd\rho+  \mmfrac{1}{\hat g} \, f\dd y \,,
\end{equation}
where $f\in\mathbb Z$ is the flux along the circle~\cite{Grimm:2015zea}. We find 
\begin{equation}\label{GS1}
S_3^\gs=  \frac{g^2}{12 \pi} \,\, C\int f A \wedge F -2 \, \zeta  \dd\rho \wedge F\,.
\end{equation}
We now wish to check that the variation of this action indeed cancels the variation of the three-dimensional one-loop consistent effective action, which we obtained in~\eqref{PVreg}, and whose variation was already inferred in~\eqref{var3d}.
In the same way that the four-dimensional gauge transformations induce three-dimensional gauge transformations and discrete shifts of the scalar, the variation of the four-dimensional scalar $\hat \rho$ induces variations of the three-dimensional scalar $\delta \rho = \theta \lambda$ and discrete shifts of the flux $\delta f =  n \, \theta$. Then~\eqref{GS1} varies as
\begin{equation}
\delta S_3^\gs = \frac{g^2}{12 \pi}     \, \theta C \, \int  n \, A \wedge F - 2 \,  \chi  \dd \lambda \wedge F \,,
\end{equation}
which, using the relation~\eqref{AC}, indeed cancels with~\eqref{var3d}, once again showing that the three-dimensional consistent effective action is gauge invariant if and only if four-dimensional anomalies are canceled.

In three dimensions, a scalar with a shift symmetry is dual to a vector, so one may dualize the scalar $\rho$ into a vector~$\tilde A$. This dualization is in fact crucial for matching the three-dimensional theory with the reduction of M-theory on a fourfold, as is explained in section~\ref{sec:M-theory}. On top of that, it also makes the anomaly condition more manifest: in that frame, the CS coefficients are integers if and only if~\eqref{AC} is satisfied. After dualization, one finds the three-dimensional effective action contains the CS terms
\begin{align}\label{CSdual}
 S^{\rm eff}_3\supset \frac{g^2}{4\pi} \int  \left [ q^2 \floor{q\, \zeta} + \mmfrac 12 \,  q^2 + \mmfrac 13 \,  C f   -  \mmfrac 23   \, \left (\theta \, C+ q^3\right )   \zeta \, \right ] A\wedge F + \mmfrac 13 \,\theta \, C\, A\wedge \tilde F \,.
\end{align}
We see that the action contains field-dependent CS terms which cancel if and only if anomalies are canceled.

In the case of an anomaly free theory, the CS coupling is (for the case of multiple fermions)
\begin{align}\label{asdasdasd}
\Theta^\af = \sum_a q_a^2 \left (\floor{q_a \vz} + \mmfrac 12  \,  \right ) + \mmfrac 13 \, f \, C \,,
\end{align}
where $f$ is an integer. At this point, it might not be clear at first sight that~\eqref{asdasdasd} is an integer, since $\sum_{\f } q_\f^3$ is not zero so the argument given below~\eqref{theta_af} does not apply anymore. However, since the scalar is periodic with period $1/\hat g$, $C$ must be a multiple of $3$ when the four-dimensional manifold is spin and a multiple of 6 in case it is spin$^c$.\footnote{This is because the last term in~\eqref{GSact} must be compatible with $\hat\rho\to\hat\rho+1/\hat g$.}
In the former case, one cannot say anything about $q^2$, consistent with the fact that on spin manifolds one can define CS theories at half-integer level~\cite{Dijkgraaf:1989pz}. In the latter case, the anomaly cancellation condition implies that $\sum_{a}q_a^3= 0 \mod 6 $, so using the same argument as below eq.~\eqref{theta_af} we conclude that the CS levels are (half)-integers, as required for consistency.

% % % % % % % % % % % % % % % % % % % % % % % % % % % % % % % % % % % % % % % % % %
\section{4D Chiral anomaly on a circle, including gravity} \label{sec:anom_circle_grav}
% % % % % % % % % % % % % % % % % % % % % % % % % % % % % % % % % % % % % % % % % % % % %

% % % % % % % % % % % % % % % % % % % % % % % % % % % % % % % % % % % % % % % % % %
\subsection{Classical compactification}

In section~\ref{sec:aom_circle} we assumed the space to be flat. 
In the present section, we consider the case in which the fermion couples to a background metric and take into account the KK vector that appears once we compactify on a circle. The starting point is the four-dimensional action
\begin{align}\label{4dactG}
S_4 = \int  \bar {\hat \psi} \,\hat \gamma^\m \left( i \hat \partial_\m + \mmfrac i2 \, \hat \omega_\mu + q \hat g \hat{A}_\mu \right) P_L \, \hat \psi \star 1 \,,
\end{align}
where $\hat \omega = \hat \omega_{\bm \bn} \hat \gamma^{\bm \bn }$ is the spin connection (barred indices indicate flat indices). 
In addition to the previous pure gauge anomaly~\eqref{var}, this theory also has a mixed gauge-gravity anomaly, which involves a triangle diagram where two of the external legs are gravitons. The variation of the 1PI effective action is then
\begin{equation}\label{mixed_an}
\delta S_4^{\textsc {1PI}} =  \frac{\hat g \, q}{24 \pi^2} \int  \, \hat \lambda \left(\,\hat g^2  q^2 \, \hat F \wedge \hat F +  {\rm Tr} \, \hat {\c R }\wedge \hat {\c R} \, \right)\,. 
\end{equation}
where $\hat {\c R} = \d\hat \omega + \hat \omega \wedge \hat \omega$ is the curvature 2-form.
In order to be anomaly free, a theory with multiple fermions coupled to gravity and with charges $q_a$ under a $U(1)$ must therefore satisfy \begin{subequations} \label{ACG}
\vspace{-3mm}
\begin{align}
\sum_\f q_a^3 &= 0\,, \label{cubic}\\
\sum_{\f} q_a &=0 \,. \label{mixed}
\end{align}
\end{subequations}

We proceed to reduce this theory on a circle with metric
\begin{equation}
\d \hat s^2 = \d s^2 - r^2 (\d y - A^0)^2 \,,
\end{equation}
where $A^0$ is the KK vector or graviphoton with field strength $F^0=\d A^0$.
A choice of vielbein for this metric is 
\begin{equation}
\hat e^\ba = e^\ba, \qquad \hat e^\bt = r \, (\d y - A^0)\,,
\end{equation}
so the $\g$-matrices with world indices are given in terms of the flat $\g$-matrices by
\begin{equation}
\hat \g^a = e^a_\ba \g^\ba, \qquad \hat \g^3 =   \g^\ba A^0_\ba + \mtfrac 1r \g^\bt \,.
\end{equation}
For the spin connection, one finds that
\begin{align}
\begin{split}
\hat \omega_{\ba \bb} & 
= \omega_{\ba \bb} + \mfrac{1}{2} r^2 F^0_{\ba \bb} \, (\d y - A^0) , \\
\hat \omega_{\ba \bt} & 
=\mmfrac 12  r F^0_{\ba b} \d x^b -  \partial_\ba r \, (\d y -A^0)  \,,  
\end{split}
\end{align}
where $\partial_\ba = e^a_\ba \, \partial_a$. 
Concerning the Ansätze of the fields, we modify the Ansatz for $\hat A$ to 
\begin{equation}
\hat A (x,y) = \mfrac{1}{\sqrt {2 \pi \,  r}} \, A(x) +  \mmfrac 1{\hat g}\, \zeta(x) (\dd y -A^0)+ \ldots \label{ansatzAG},
\end{equation}
and keep the one for $\hat \psi $ unchanged.

The reduction of the action~\eqref{4dactG} is\footnote{We do not include couplings of the fermions to $\partial_a r$, since they are irrelevant for our analysis.}
\begin{equation}\label{3dactG}
S_3=\sum_{n \in \Z}  \int  \bar \psi_n \, \Big [\gamma^a \big(
 i  \partial_a +\mmfrac i2 \,\omega_a  \!+ q g A_a 
+ n A^0_a - \mfrac r{8} \epsilon_{abc} F^{0bc}\big)
+ \mtfrac 1r \g^\bt \left( n + q \, \zeta \right)
\Big ]P_L \, \psi_n\star 1 \,,
\end{equation}
As before, the mass of the $n$-th KK mode is $m_n = \frac 1r (n+ q \vz)$ in the CB.
We see that the $n$-th KK mode has charge $n$ under the graviphoton, as expected. In addition, we find a Pauli coupling $\sim F^0 \bar \psi\psi$ between the fermions and $A^0$.

One could proceed as in section~\ref{sec:aom_circle} and reduce the gauge variation of the 1PI action~\eqref{var} including the new Ansatz for $\hat A$, and get the expected three-dimensional variation, but by now the logic is clear and we proceed directly to integrating out the KK-tower of fermions in three dimensions.

% % % % % % % % % % % % % % % % % % % % % % % % % % % % % % % % % % % % % % %
\subsection{Integrating out the Kaluza-Klein modes}

Integrating out the massive fermions in three dimensions induces CS couplings and when the coupling to gravity is taken into account there can be three types of couplings, namely
\begin{equation}\label{CSlagr}
\c L^{\cs} = 
\frac{g^2}{4\pi}\, \Theta \, A \wedge F +
 2 \, \frac{g}{4\pi}\, \Theta_0 \, A \wedge F^0 + 
 \frac{1}{4\pi} \, \Theta_{00} \, A^0 \wedge F^0 \,.
\end{equation}
Using the formula~\eqref{int_ferm} with $q=n$ for $A^0$, we find the following expressions for the $\Theta$ coefficients\footnote{One might be worried that the contribution~\eqref{int_ferm} of a single massive mode gets modified by the Pauli coupling in~\eqref{4dactG}. This however does not happen, as discussed in appendix~\ref{ap:3dreg}.} 
\begin{align} \label{sums}
\begin{split}
\Theta & = \mmfrac 12  \, q^2 \sum_{n \in \Z} \sign \left (n + q \vz  \right ) \,, \\
\Theta_0 &= \mmfrac 12  \, q  \sum_{n \in \Z}  n \sign \left (n + q \vz  \right ) \,, \\
\Theta_{00}& = \mmfrac 12  \,  \sum_{n \in \Z} n^2\sign \left (n + q\vz \right ) \,,
\end{split}
\end{align}
The zeta-function regularization of these sums is (see appendix~\ref{ap:Sums})
\begin{align}  \label{summedG}
\begin{split}
\Theta^\reg& =  q^2 \left( \mmfrac 12  + \floor{q\,\vz} -  q\,\vz \right), \\[3pt]
\Theta^\reg_0 &= q \left( - \mmfrac 1{12} - \mmfrac 12  \floor{q\,\vz} (\floor{q\,\vz}+1 ) + \mmfrac 12 \, q^2\vz^2
 \right)  , \\[3pt]
\Theta^\reg_{00}& = \mmfrac{1}{6} \floor{q\,\vz}  (\floor{q\,\vz} + 1) (2\floor{q\,\vz} +1)- \mmfrac 13 \, q^3\vz^3\,. 
\end{split}
\end{align}
We can also obtain these results by regulating the one-loop diagrams that produce eqs.~\eqref{sums}, using Pauli-Villars regularization in three dimensions. We then have to introduce a set of three PV regulators for each KK-mode, as shown in appendix~\ref{ap:3dreg}.  
We see once again that for an anomalous theory, the CS coefficients depend continuously on the Coulomb branch parameter $\vz$. This dependence drops when anomalies are canceled.

Let us look now at the transformations of these $\Theta$'s under large gauge transformations of the four-dimensional vector, $\delta \hat A = \frac{1}{\hat g} \, n \d y$. From~\eqref{ansatzAG}, we see that such a transformation acts on the three-dimensional fields as
\begin{align}\label{qweqwe}
\delta A=\mfrac{1}{g}A^0\,,\qquad \delta A^0=0\,,\qquad \delta \zeta = n\,.
\end{align}
Thus, the CS coefficients shift as
\begin{align}\label{shifts}
\begin{split}
\Theta & \to \Theta \,,\\
\Theta_0 & \to \Theta_{0} - \mfrac{\raisebox{-1.5pt}{$n$}}{\raisebox{1.5pt}{$g$}} \, \Theta \,, \\
\Theta_{00} & \to \Theta_{00} - 2  \, \mfrac{\raisebox{-1.5pt}{$n$}}{\raisebox{1.5pt}{$g$}} \, \Theta_{0} +  \mfrac{\raisebox{-1.5pt}{$n^2$}}{g^2} \, \Theta \,.
\end{split}
\end{align}
One can check that under the transformations~\eqref{qweqwe}, together with these shifts~\eqref{shifts}, the CS Lagrangian~\eqref{CSlagr} is invariant. For an anomalous theory, one does not expect the three-dimensional Lagrangian to be invariant under gauge transformations and, in particular, it should not be invariant under these shifts. However, the CS terms~\eqref{summedG} do make the three-dimensional action invariant, similarly to the discussion below eq.~\eqref{summed}. In order to find the CS coefficients that correctly reproduce the anomalous variation under gauge transformations, we have to regularize the theory using the (non gauge-invariant) PV regulator that comes from four dimensions, which we discuss in the following.

% % % % % % % % % % % % % % % % % % % % % % % % % % % % % % % % % % % % % % % % % % % % % %
\subsection{Integrating out the Kaluza-Klein modes, preserving the 4D symmetries}
\label{sec:PVregG}

In this case we use the same regulator as in~\eqref{S4PVs}, but with the coupling to gravity included,
\begin{equation}\label{S4PVG}
S_4^\pv =  \int \sums  \,\bar{\hat{\psi}}_s \left [ \hat \gamma^\m \left( i \hat \partial_\m   + \mmfrac i2 \,\hat \omega_\mu P_L + q \hat g \hat{ A }_\m \,  P_L  \right) +M_s\right ]\hat \psi_s \, \, \hat \star \, 1 \,.
\end{equation}
Reducing this action on a circle we find 
 \begin{equation}\label{3DPVG}
 \begin{split}
  S_3=\sum_{n \in \Z}  \int \sums  \bar\psi_{n,s} \, \Big \{ \gamma^a \!\left( i  \partial_a   + \mmfrac i2 \, \omega_a P_L \!+ q g A_a P_L + n A^0_a - \mfrac r{8} \epsilon_{abc} F^{0bc} P_L \right) \quad \\
+ \mtfrac 1r \gamma^\bt (n + q \, \zeta P_L) 
+ M_s  \Big \} \, \psi_{n,s} \star 1 \,.
 \end{split}
 \end{equation}
  \begin{figure}
  \centering 
  \begin{subfigure}{0.4\textwidth}
  \centering
  \includegraphics[scale=1.15]{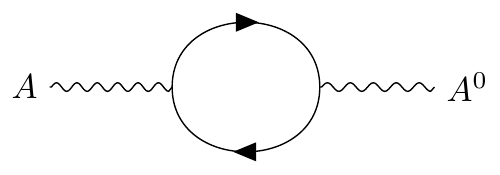}
  \end{subfigure}
  \hspace{3mm}
  \begin{subfigure}{0.4\textwidth}
  \centering
  \includegraphics[scale=1.15]{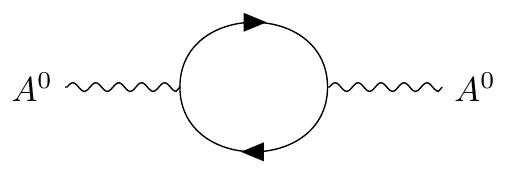}
  \end{subfigure}
  \caption{One-loop diagrams contributing to the field independent CS terms involving the KK photon $A^0$ in three dimensions.}
  \label{fig:oneloopCSG}
  \end{figure}
  
From this action, one can extract the Feynman rules and compute the diagrams of figure~\ref{fig:oneloopCSG}, which is done in detail in appendix~\ref{ap:4dreg}. We note that the Pauli coupling $\sim F^0 \bar \psi_n\psi_n$ has to be included in the computation of the diagram.\footnote{This only contributes to the divergent terms which precisely cancel the divergence coming from the minimal coupling $n A^0 \bar \psi_n\psi_n$, making the final result finite. A similar observation was also made in~\cite{Bonetti:2013ela}.} 
The resulting one-loop CS coefficients are indeed different from those in~\eqref{summedG}; they read
\begin{align} \label{PVregG}
\begin{split}
\Theta^\regb& =  q^2 \left( \mmfrac 12  + \floor{q\,\zeta} -\mmfrac 23 \,  q\,\zeta \right), \\[3pt]
\Theta^\regb_0 &= q \left( - \mmfrac 1{12} - \mmfrac 12  \floor{q\,\zeta} (\floor{q\,\zeta}+1 ) + \mmfrac 16 \,q^2\zeta^2
 \right)  , \\[3pt]
\Theta^\regb_{00}& = \mmfrac{1}{6} \floor{q\,\zeta}  (\floor{q\,\zeta} + 1) (2\floor{q\,\zeta} +1)\,. 
\end{split}
\end{align}
Once again, we see that the regularization procedure we used gives CS terms which are field-dependent, making the effective action not invariant under small gauge transformations.
In addition, these coefficients do not shift like~\eqref{shifts}, and the CS Lagrangian~\eqref{CSlagr} is not invariant under~\eqref{qweqwe}, as expected for an anomalous theory. One can check that its variation matches with the circle reduction of the variation~\eqref{mixed_an}, using the Ansatz~\eqref{ansatzAG}. This tells us that the coefficients~\eqref{PVregG} lead to the consistent effective action. 

\subsection{Anomaly inflow}
The relation between covariant and consistent effective actions, through anomaly inflow, as was briefly explained in section~\ref{sec:AI}, can be generalized to curved backgrounds~\cite{Jensen:2013kka,DiPietro:2014bca}. In this case, eq.~\eqref{AI}  gets extended to 
\begin{equation} \label{AIG}
S_5
= -\frac{g^2}{12 \pi} \, q^3 \, \int_{\mathcal M_3 } \vz \, A \wedge F - 2 \vz^2 A \wedge F^0 + \vz^3 A^0 \wedge F^0\,,
\end{equation}
which is precisely the difference between~\eqref{summedG} and~\eqref{PVregG}, meaning that the relation~\eqref{thetasrel} also holds for $\Theta_0$ and $\Theta_{00}$.

% % % % % % % % % % % % % % % % % % % % % % % % % % % % % % % % % % % % % % % % % % % % %
\subsection{Chern-Simons coefficients for an anomaly-free theory}
As already noted above, the field-dependent terms in the $\Theta$'s are all proportional to the anomaly coefficient $q^3$, so in a theory canceling the pure gauge anomaly, i.e~when $\sum_{\f} q_\f^3 = 0$ (recall that $a$ labels the different fermions), we find 
\begin{align} \label{theta_afG}
\begin{split}
\Theta^{\af}& =  \sum_a q_\f^2 \left( \mmfrac 12  + \floor{q_a \vz } \right), \\
\Theta^\af_0 &= \sum_a q_\f  \left( - \mmfrac 1{12} - \mmfrac 12  \floor{q_a \vz } (\floor{q_a \vz }+1 )
 \right)  ,\\
\Theta^\af_{00}& = \sum_a \mmfrac{1}{6} \floor{q_a \vz }  (\floor{q_a \vz }+ 1) (2\floor{q_a \vz } +1) \,. 
\end{split}
\end{align}
These CS terms were already found in \cite{Grimm:2015zea} and several important aspects were previously discussed in~\cite{Golkar:2012kb, Cvetic:2012xn,Jensen:2013kka,Jensen:2013rga,DiPietro:2014bca,DiPietro:2016ond}. 
The CS coefficients~\eqref{theta_afG} are discrete and are also in fact integers, as follows.
We showed in section~\ref{sec:aom_circle} that $\frac 12 \sum_{\f} q_a^2 \in \Z $. 
However, the term $-\frac{1}{12} \sum_{\f} q_a$ is in general half integer\footnote{Notice that $q^3-q = q(q-1)(q+1) \equiv 0 \mod 6$, such that $\frac{1}{12}\sum_a q_a\in \frac 12 \Z$ if $\sum_a q_a^3 =0$.}  so seems that $\Theta^\af_0$ might not be integer.
However, for a four-dimensional theory (coupled to gravity) to be consistent, it needs to cancel both the cubic Abelian anomaly~\eqref{cubic} as well as the mixed gauge-gravitational anomaly~\eqref{mixed}. 
In that case, this term vanishes and the three $\Theta$'s in~\eqref{theta_afG} are integers. In contrast, in a theory that does not cancel the mixed anomaly $\Theta_{0}$ could be half integer, signaling an inconsistency if the manifold is spin$^c$. However this is not guaranteed, it could happen that $\sum_a q_a \equiv 0 \mod 12 $, thus the 3D gauge CS terms being consistent, whilst having an anomalous four-dimensional theory. Notice that here we are only computing CS terms for the vectors $A$ and $A^0$ but, in principle, one could also compute the three-dimensional gravitational CS term, introduced below in~\eqref{CS-form}. We expect this to be field-dependent when the mixed gauge-gravitational anomaly is not canceled, signaling the inconsistency in four-dimensional.

% % % % % % % % % % % % % % % % % % % % % % % % % % % % % % % % % %
\subsection{Green-Schwarz mechanism}
In this section we include the possibility of having Green-Schwarz terms in four dimensions, both for the pure gauge anomaly and the mixed one. That is, we add to the action~\eqref{4dactG} the non gauge invariant term
\begin{equation}\label{GSactG}
S_4^\gs= 
\frac{\hat g}{24 \pi^2}  \int  \hat\rho \ \left( \hat g^2 C \, \hat F\wedge\hat F + a \,   {\rm Tr} \, \hat {\c R} \wedge \hat {\c R} \right)\,,
\end{equation}
whose variation cancels with~\eqref{mixed_an} if and only if the Green-Schwarz parameters $C$ and $a$ satisfy the conditions
\vspace{-3mm}
\begin{align}
 q^3 + \theta \, C & =0 \label{cubic-GS}\,,\\
q +\theta \, a & =0  \label{mixed-GS}\,.
\end{align}
Upon reduction on a circle, only the term $\sim \hat F\wedge \hat F$ contributes to the CS coupling. 
Using the Ansatz~\eqref{ansatzAG} for $\hat A$ and~\eqref{ansatz_rho} for $\hat \rho$, we find the following CS piece (after dualizing $\rho$ into a vector $\tilde A$)
\begin{align}
\begin{split}
\!\!\!\!\!\! S_3^\cs  = \frac{g^2}{4\pi}\int      \left( - \mmfrac 23 \, \theta \,  \zeta  + \mmfrac 13\, f \, \right ) C  A \wedge F + \mmfrac 13\, \theta  \, C \, A \wedge \tilde F \qquad\qquad\\
+ \frac{g}{4\pi} \int \mmfrac 13 \, \theta \, C \, \zeta ^2 A \wedge F^0 - \mmfrac 13 \,f \, C\,  A^0 \wedge \tilde F .
\end{split}
\end{align}
The two terms that depend on $\zeta$ combine with the consistent one-loop terms found in~\eqref{PVregG} so the full CS terms\footnote{Involving only $A$ and $A^0$, not $\tilde A$.} in the three-dimensional effective action is
\begin{equation}\label{CS-GS}
\begin{aligned} 
\Theta^{\textsc{gs}}& =  \mmfrac 12 \, q^2  + q^2 \floor{q\, \zeta } + \mmfrac 13 \, f \, C -\mmfrac 23 \, (q^3 +  \theta \, C) \, \zeta \, ,\\[3pt]
\Theta^{\textsc{gs}}_0 &= - \mfrac q{12} - \mfrac q2 \floor{q\, \zeta  } (\floor{q\, \zeta  }+1 ) + \mmfrac 16 \, (q^3 +  \theta \, C) \zeta^2 \, , \\[3pt]
\Theta^{\textsc{gs}}_{00}& = \mfrac{1}{6} \floor{q\, \zeta  }  (\floor{q\, \zeta  } + 1) (2\floor{q\, \zeta  } +1)\,.
\end{aligned}
\end{equation}
Again we find that the field-dependent contribution to the terms are multiplied by the anomaly~\eqref{cubic-GS}, meaning that they vanish if and only if anomalies cancel. In order to see that the remaining CS coefficients are integers, we again notice that Green-Schwarz parameters have to satisfy $C \equiv 0 \mod 6$ and $a \equiv 0 \mod 12$ when the manifold is spin$^c$ and $C \equiv 0 \mod 3$ and $a \equiv 0 \mod 6$ when the manifold is spin. 
This implies that the CS levels are integers in the former case and half-integer in the latter. 

% % % % % % % % % % % % % % % % % % % % % % % % % % % % % % % % % % % % % % % % % % % % % % %
\subsection{General case}
\label{gen_case}
In this section we consider the generalization of the previous analysis to the case with multiple fermions charged under multiple abelian gauge fields. In order to also include non-abelian gauge theories, we note that for such theories, once we compactify on a circle and go to a generic point in the Coulomb branch, the gauge group gets broken to its Cartan subgroup. Anomaly cancellation can then be inferred by focusing on the abelian subsector and following the analysis below.

We thus proceed to generalize the actions~\eqref{4dactG} and~\eqref{GSactG} 
to the case with multiple fermions $\Psi_a, \ a=1,\ldots,N_f$, with charges $q_{a,i}$ under multiple $U(1)$ gauge fields $A^i, \ i = 1, \ldots,N_{U(1)}$.
and multiple scalars $\rho_\a, \ \a =1,\ldots,N_\rho$, with covariant derivatives $\hat D \hat \rho_\a = \d \hat \rho_\a - \theta_{i\a} \hat A^i$.
 Having several $U(1)$'s, one can include generalized CS terms~\cite{Anastasopoulos:2006cz}, which are of the form $A^i \wedge A^j \wedge F^k$ (these vanish identically when only a single $U(1)$ is present). 
Similar topological terms, involving the gravitational CS-form
\begin{equation}\label{CS-form}
\hat \Omega = \tr{\hat \omega \wedge \d \hat \omega + \mmfrac 23 \, \hat \omega \wedge \hat \omega \wedge \hat \omega}
\end{equation}
can additionally be considered.
The four-dimensional action that we consider in this section is then
\begin{equation}
\begin{split}
S_4 = \int \sum_a \bar {\hat \Psi}_a  \hat \gamma^\m \left( i \partial_\m \!+\! \mfrac i 2 \hat \omega_\mu  + q_{a, i}\,  \hat g \hat{A}^i_{ \mu} \right) \! P_L \hat \Psi_a \star \! \, 1 \hspace*{5.5cm}
 \\\,\, +\frac{\hat g^3 }{24\pi^2}  
 \big ( C^\a_{ij} \,  \hat \rho_\a  \, \hat F^i \! \wedge\hat F^j- E_{ijk} \hat A^i \wedge \hat A^j \wedge \hat F^k \big )
+\frac{\hat g }{24\pi^2}   \big (a^\a \hat \rho_\a \, {\rm tr} \, \hat {\c R} \! \wedge \!\hat{ \c R} 
-  \, b_i \hat A^i \wedge  \hat \Omega \big)\,. \!\!\!
\end{split}
\end{equation}
The one-loop variation of this action is 
\begin{equation}
\delta S_4^\oneloop = \frac{\hat g}{24\pi^2}\sum_a \int\hat \lambda^i  \left (\hat g^2  q_{a,i} q_{a,j} q_{a,k} \hat F^j \wedge \hat F^k + q_{a,i}  \,{\rm tr} \hat {\c R} \wedge \hat{ \c R} \right)\,,
\end{equation}
which is the generalization of~\eqref{mixed_an}. Its classical variation, on the other hand, reads 
\begin{equation}
\delta S_4^{\rm class} = \frac{\hat g }{24\pi^2} \int\hat \lambda^i  \left[\hat g^2 \left(\theta_{i\a} C^\a_{jk} + E_{ijk} + E_{ikj}\right) \hat F^j \wedge \hat F^k + \left( \theta_{i\a} a^\a + b_i \right)  \,{\rm tr} \hat {\c R} \wedge \hat{ \c R} \right]\,,
\end{equation}
such that anomaly cancellation conditions take the form
\begin{subequations}
\begin{align} \label{AC2-G}
\sum_a q_{a,i} q_{a,j} q_{a,k}+\theta _{i\a} C^\a_{jk} + E_{ijk} + E_{ikj}  &=0\,, \\
\sum_a q_{a,i} + \theta _{i\a} a^\a + b_i &=0\,.
\end{align}
\end{subequations}

We now want to compactify this action on a circle, go to the Coulomb branch, and integrate out the massive fermions, again regularizing while preserving four-dimensional Lorentz invariance, as is done earlier in the text. The resulting one-loop CS couplings are
\begin{align} \label{gen3dCSoneloopG}
\begin{split}
\Theta^\oneloop_{ij} & =  \sum_a q_{a,i} q_{a,j} \left ( \mmfrac 12   + \floor{q_{a,k}\zeta^k } - \mmfrac 23 \,\zeta^k \right) \,, \\
\Theta^\oneloop_{0i} &=  \sum_a q_{a,i} \left( - \mmfrac 1{12} - \mmfrac 12  \floor{q_{a,j}\zeta^j  } (\floor{q_{a,k}\zeta^k }+1 ) + \mmfrac 16 \,  q_{a,j} q_{a,k}\zeta^j \zeta^k
 \right)  , \\
\Theta^\oneloop_{00}& = \sum_a \mfrac{1}{6} \, \floor{q_{a,i}\zeta^i  }  (\floor{q_{a,j}\zeta^j  } + 1) (2\floor{q_{a,k}\zeta^k } +1) \,.
\end{split}
\end{align}
The Green-Schwarz and generalized CS terms also yield three-dimensional CS couplings upon circle reduction, and dualization of the scalars $\rho_\a$ into vectors $\tilde A^\a$. 
The resulting classical CS coefficients are
\begin{align} \label{gen3dCSclassG}
\begin{split}
\Theta^{\rm class}_{ij} & = \mmfrac 13 \,  f_\a C^\a_{ij} - \mmfrac 23 \left( \theta _{i\a} C^\a_{jk} + E_{ijk} + E_{ikj} \right) \zeta^k\,,  \\[3pt]
\Theta^{\rm class}_{0i} &= \mmfrac 16 \left( \theta _{i\a} C^\a_{jk} + E_{ijk} + E_{ikj} \right)\zeta^j\zeta^k \,,\\[3pt]
\Theta^{\rm class}_{00}& = 0  \,.
\end{split}
\end{align}
Combining~\eqref{gen3dCSoneloopG} and~\eqref{gen3dCSclassG}, we find that the field-dependent CS coefficients are
\begin{align}
\begin{split}
\Theta^{\rm \text{field-dep}}_{ij} &= -\mmfrac 23 \, \Big( \!\sum_a q_{a,i} q_{a,j} q_{a,k} + \theta _{i\a} C^\a_{jk} + E_{ijk} + E_{ikj} \!\Big) \, \zeta^k  \,,\\
\Theta^{\rm \text{field-dep}}_{0i} &= \mmfrac 16  \, \Big( \!\sum_a q_{a,i} q_{a,j} q_{a,k} + \theta _{i\a} C^\a_{jk} + E_{ijk} + E_{ikj}\!\Big) \, \zeta^j\zeta^k\,,
\end{split}
\end{align}
so they are proportional to the anomaly~\eqref{AC2-G}, and therefore vanish in an anomaly free theory. 
We again find that the remaining CS
coefficients are (half)-integers.

%%%%%%%%%%%%%%%%%%%%%%%%%%%%%%%%%%%%%%%%%%%%%%%%
\section{\boldmath M-theory reduction with $G_4$-flux} \label{sec:M-theory}
%%%%%%%%%%%%%%%%%%%%%%%%%%%%%%%%%%%%%%%%%%%%%%%%

In order to obtain effective actions of F-theory, we use the definition of F-theory as a special limit of M-theory. Note that M-theory on a Calabi-Yau fourfold with flux yields a three-dimensional $\c N = 2 $ effective theory~\cite{Becker:1996gj}. Considering the fourfold to be two-torus fibered then a four-dimensional effective action is obtained in the limit of vanishing torus fiber. However is not known how to perform this singular limit, since non-perturbative states of M-theory become relevant.
Therefore, we rather infer information about the four-dimensional theory by comparing the three-dimensional $\c N = 2 $ theory obtained from M-theory with the circle compactification of a generic four-dimensional $\c N =1$ theory~\cite{Grimm:2010ks,Grimm:2011fx}. In the process of compactifying the four-dimensional theory on a circle, going into the Coulomb branch and computing the Wilsonian effective action, one might loose information. Nevertheless, one can compare this three-dimensional effective action with the one obtained by standard dimensional reduction of eleven-dimensional supergravity on a smoothed Calabi-Yau fourfold with flux~\cite{Haack:1999zv,Haack:2001jz}.

As we have seen in previous sections, an anomalous theory in four dimensions, once compactified on a circle, generates field-dependent CS couplings, which are not gauge invariant. In fact, these field-dependence drops if and only if anomalies cancel in the four-dimensional theory. So, as far as anomalies are concerned, the information is not lost upon circle compactification. This means that, instead of worrying about proving anomaly cancellation in four dimensions, we may use the indirect and general approach of dimensional reduction of eleven-dimensional supergravity on a smooth Calabi-Yau fourfold with flux. As we will recall below, no such field-dependent CS terms are found in this reduction, such that the four-dimensional theory obtained from compactification of F-theory is free from local anomalies. This proof, of course, requires the existence of a \emph{smooth} Calabi-Yau fourfold with background flux which describes the F-theory setting in the singular limit, which is not always the case, see for instance~\cite{10.2307/2044659,Hanany:2000fq,Donagi:2003hh,Cecotti:2010bp,Anderson:2013rka,Collinucci:2014taa,Arras:2016evy}.

In the following we briefly review the dimensional reduction of eleven-dimensional supergravity on a Calabi-Yau fourfold $Y_4$ including $G_4$-flux~\cite{Haack:1999zv,Haack:2001jz}. 
The bosonic action is given by~\cite{Cremmer:1978km}
\begin{equation}\label{11Dsugra}
S_{11} = \int\left( \mmfrac 12 \,  R_{11} \star_{11} 1 - \mmfrac 14 \d \C \wedge \star_{11} \d \C - \mmfrac{1}{12} \, \C \wedge \d \C \wedge \d \C\right)\,.
\end{equation}
Since we are interested in the CS terms that appear in the three-dimensional effective theory,
it is enough to consider the dimensional reduction of the eleven-dimensional CS term which involves only the three-form $C_3$. We may consider the following reduction Ansatz
\begin{align}
C_3=\langle C_3\rangle + A^\Sigma\wedge \omega_\Sigma+N_{\mathcal A}\Psi^{\mathcal A}+\bar N_{\mathcal A}\bar \Psi^{\mathcal A}
\end{align}
where $\omega_\Sigma$ is a basis of harmonic two-forms in $Y^4$ and $\Psi^{\mathcal A}$ is a basis of $(2,1)$-forms. Here $A^\Sigma$ are three-dimensional vectors and $N_{\mathcal A}$ are complex scalars. The background value of $C_3$ is such that $\langle G_4\rangle=\langle \d C_3\rangle$ is an appropriately quantized four-form in $Y_4$~\cite{Witten:1996md}. Using this Ansatz, one can readily reduce the last term in~\eqref{11Dsugra} which gives the topological terms
\begin{align}\label{ertert}
S_3=-\nfrac{1}{4} \int \Big [\Theta_{\Sigma\Lambda} A^\Sigma\wedge F^\Lambda -i \, d_\S{}^{\c A \c B} F^\S \wedge\big( N_{\c A} D \bar N_{\c B} - N_{\c B} D \bar N_{\c A} \big) \Big ] \,,
\end{align} 
with
\vspace*{-4mm}
\begin{align}\label{int_nb}
\Theta_{\Sigma\Lambda} &= \int_{Y_4} \omega_\Sigma \wedge \omega_\Lambda \wedge \langle G_4\rangle\,,\\[2pt]
d_\S{}^{\c A \c B}& = i \int_{Y_4} \omega_\S \wedge \Psi^{\c A} \wedge \bar \Psi^{\c B} \,.
\end{align}
The CS terms are the first term in~\eqref{ertert}, which are clearly field-independent, since the $\Theta_{\S \Lambda}$ are intersection numbers. Furthermore, the quantization of the $G_4$-flux translates into the quantization of the Chern-Simons.  
The derivatives $DN_{\c A}$ in~\eqref{ertert} contain the coupling to complex structure moduli but no gaugings by the vectors $A^\S$ (see e.g.~\cite{Corvilain:2016kwe}). Therefore the second term does not include further CS couplings. 

In general, the three-dimensional action~\eqref{ertert} is not in the correct frame to match with the circle reduction of a general 4D $\c N=1$ supergravity theory: some of the scalars correspond to vectors in 4D, and vice-versa. Thus performing a field dualisation in 3D is required to lift to 4D (see~\cite{Corvilain:2016kwe} for a more detailed discussion).
However, since the action~\eqref{ertert} is completely gauge invariant, it will remain so after dualization.

This corresponds, therefore, to the circle compactification of an anomaly-free theory, as expected. This concludes the proof that four-dimensional F-theory compactifications that can be obtained as a limit of M-theory on a smooth Calabi-Yau fourfold with background flux have no local anomalies.

Let us close this section with some remarks on the implications and extensions of our result. 
Firstly, we note that, under our assumptions, the geometric relations derived from assuming gauge anomaly cancellation found in~\cite{Cvetic:2012xn,Bies:2017abs} are indeed satisfied. 
Secondly, our strategy might also be used to translate flux quantization conditions in M-theory to flux quantization conditions in type IIB string theory with seven-branes, which has been previously discussed from a different perspective in~\cite{Collinucci:2010gz,Collinucci:2012as}.
Thirdly, it is important to stress that we did not discuss the cancellation of anomalies of geometrically massive $U(1)$'s. They can be included~\cite{Grimm:2011tb,Braun:2011zm,Braun:2014nva,Grimm:2015ona} and lead to a gauging of the complex fields $N_{\c A}$ in the effective Lagrangian~\eqref{ertert}.  It would be interesting to extend our discussion to this case.

%%%%%%%%%%%%%%%%%%%%%%%%%%%%%%%%%%%%%%%%%%%%%%%%
\section{Conclusion and Outlook} \label{sec:conclusions}
%%%%%%%%%%%%%%%%%%%%%%%%%%%%%%%%%%%%%%%%%%%%%%%%
The cancellation of local anomalies is of crucial importance to ensure the quantum consistency of a gauge theory. In four-dimensional gauge theories such local anomalies arise from chiral fermions and lead to an associated chiral anomaly via one-loop diagrams. In contrast, it is well-known that there are no local anomalies in odd-dimensional theories. This immediately raises the question: Where is the information about the four-dimensional anomaly once the theory is compactified to three dimensions?

In this work, we have answered this question for a four-dimensional gauge theory with chiral fermions and the inclusion of a possible Green-Schwarz mechanism. We have shown that the four-dimensional chiral anomaly is visible after compactification in the form of the classical and one-loop Chern-Simons terms. It was hereby crucial that (1) all Kaluza-Klein modes of the chiral fermions are included in the one-loop diagrams, and that (2)  a regularization scheme is used that preserves four-dimensional Lorentz invariance. 
We have thus performed the compactification of the four-dimensional (non gauge-invariant version of) Pauli-Villars regulators.
We contrasted this with the regularization using zeta-functions, which also yields a finite result but removes part of the information about the four-dimensional anomalies; it relates to our regularization through an anomaly inflow analysis.
As a non-trivial check of our one-loop results, we have also extended the discussions to include four-dimensional Green-Schwarz couplings, generalized Chern-Simons couplings, as well as three-dimensional Chern-Simons terms involving the Kaluza-Klein vector arising from the metric. Altogether, we were thus able to show that the cancellation of four-dimensional local gauge anomalies is equivalent to demanding that appropriately regularized Chern-Simons terms are gauge invariant.

In the final part of this work we have given a first interesting application of our results. We recalled that general four-dimensional F-theory effective actions are related to an M-theory compactification after a circle compactification.
Therefore, we can reliably test the anomalies of F-theory effective actions by analyzing the M-theory effective action in three dimensions. The latter can be determined using eleven-dimensional supergravity if the F-theory Calabi-Yau geometry can be fully resolved preserving the Calabi-Yau condition and all information about the chiral spectrum is captured by a four-form flux on this resolved space. It is then straightforward to check that the three-dimensional Chern-Simons terms from M-theory are gauge invariant and hence four-dimensional anomalies are always canceled for these settings. 
This implies that studying F-theory models via M-theory the consistency conditions required for gauge anomaly cancellation are implemented automatically, if a resolved Calabi-Yau geometry with a four-form flux background exists. 
We expect that this result similarly holds for F-theory effective actions in six-dimensions. More precisely, we expect that if the F-theory compactification geometry can be fully resolved preserving the Calabi-Yau property, the six-dimensional theory is anomaly free. 
In general, we believe that our results and the developed perspective will help us to deepen our understanding of the M-theory to F-theory limit and eventually allow us to investigate other general properties of F-theory effective actions in various dimensions.      

There are several interesting open problems left for future research. Firstly, it is interesting to extend this analysis to other dimensions and to other compactification spaces. A natural next step is a generalization to circle compactifications of six-dimensional theories. In this case, it will be crucial to also include tensor modes in the five-dimensional one-loop computations of the Chern-Simons terms~\cite{Bonetti:2013ela}. 
While the general expressions for anomaly free theories have been given in~\cite{Grimm:2015zea}, it remains to compute the non-gauge-invariant pieces when starting with an anomalous theory, see however~\cite{DiPietro:2014bca} for an analysis using anomaly matching. Secondly, it is also important to extend our analysis to include other anomalies. For example, it is interesting to ask how global anomalies are manifest in the lower-dimensional theory. 
It is also a pressing open question to study the mixed gauge-gravitational anomaly, which will require the investigation of the gravitational Chern-Simon coupling. We hope to return to this in future work.

\subsection*{Acknowledgments}
		
We are grateful to I\~naki Garc\'ia-Etxebarria, Andreas Kapfer, Zohar Komargodski, Miguel Montero and Irene Valenzuela for useful discussions and to Markus Dierigl and Kilian Mayer for carefully reading and commenting this document. This work was partly supported by a grant of the Max Planck Society.

\appendix

%%%%%%%%%%%%%%%%%%%%%%%%%%%%%%%%%%%%%%%%%%%%%%%%
\section{Regularization of the sums} \label{ap:Sums}
%%%%%%%%%%%%%%%%%%%%%%%%%%%%%%%%%%%%%%%%%%%%%%%%

In this appendix, we regularize the sums~\eqref{sums} (the first of which already appears in~\eqref{sum})
\begin{align}\label{sums-ap}
\begin{split}
S_1(\mu) &= \mmfrac 12  \sum_{n\in \Z}  \sgn (n+\m)\,,\\
S_2(\mu) &= \mmfrac 12  \sum_{n\in \Z}  n \sgn (n+\m)\,,\\
S_3(\mu) &= \mmfrac 12  \sum_{n\in \Z} n^2 \sgn (n+\m)\,.
\end{split}
\end{align}
where we defined $\m=q\vz$. 
First we note that for $ \bbn \in \Z$,
\begin{align}\label{transf-sums}
\begin{split}
S_1(\m + \bbn) &= S_1(\mu) \,, \\ 
S_2(\m + \bbn) &= S_2(\mu) - \bbn S_1(\mu)\,,\\ 
S_3(\m + \bbn) &= S_3(\mu) -2 \bbn S_2(\mu) + \bbn^2 S_1(\mu) \,.
\end{split}
\end{align}
In particular, decomposing $\mu = \fr \m + \floor{\mu}$ in its fractional part $\fr \mu$ and its integral part $\floor{\mu}$, we have
\begin{align}\label{sums-fr}
\begin{split}
S_1 (\mu) &= S_1(\fr \mu) \,,\\ 
S_2 (\mu)&= S_2(\fr \mu) - \floor{\mu} S_1(\fr \mu) \,,\\ 
S_3 (\mu) &= S_3(\fr \mu) -2 \floor{\mu} S_2(\fr \mu) + \floor{\mu}^2 S_1(\fr \mu) \,,
\end{split}
\end{align}
such that we only need to compute the $S_i(\fr \m)$ for $\fr \mu \in (0,1)$.

In oder to regulate the sums in~\eqref{sums-ap}, we use a method related to the zeta function regularization, namely we first multiply the argument of the sums by $\abs{n+\fr \m}^{-s}$, 
making the sums convergent, and after evaluating them, we take the analytic continuation to $s=0$. The result can be written as 
\begin{align}\label{zeta-rest}
\begin{split}
S_1^{\reg}(\fr \m)  & = \mmfrac 12  \big [ \zeta(0,\fr \mu ) - \zeta (0,-\fr \mu ) + 1 \big ] \,,\\ 
S_2^{\reg}(\fr \m)  & = \mmfrac 12 \big [ \zeta(-1,\fr \mu ) + \zeta (-1,-\fr \mu ) - \fr \mu  \left(\zeta(0,\fr \mu ) - \zeta (0,-\fr \mu ) \right) \big ] \,,\\
S_3^{\reg}(\fr \m)  & = \mmfrac 12 \big[ \zeta(-2,\fr \mu ) - \zeta (-2,-\fr \mu ) -2 \fr \mu  \left(\zeta(-1,\fr \mu ) + \zeta (-1,-\fr \mu )\right)  %\\ &  \hspace*{7.5cm}
+ \fr \mu ^2 \left(\zeta(0,\fr \mu ) - \zeta (0,-\fr \mu) \right) \big] \,,
\end{split}
\end{align}
where $\zeta(s,x) = \sum_{n = 0}^\infty (n+x)^{-s}$ is the Hurwitz zeta function. When $s$ is a negative integer, it reduces to the Bernoulli polynomials
\begin{equation}
\zeta(-n,x) = - \frac{B_{n+1}(x)}{n+1}  \,.
\end{equation}
Using the explicit forms of those polynomials, the results~\eqref{zeta-rest} become
\begin{align}
\begin{split}
S_1^{\reg}(\fr \m)  & = \mmfrac 12  - \fr \mu \,,\\ 
S_2^{\reg}(\fr \m)  & = - \mfrac 1 {12} + \mmfrac 12 \, \fr \mu^2 \,,\\
S_3^{\reg}(\fr \m)  & = - \mmfrac 13 \, \fr \mu^3 \,.
\end{split}
\end{align}
Substituting those in eqs.~\eqref{sums-fr}, we obtain expressions of the regularized sums holding for generic $\mu \notin \Z$,
\begin{align}\label{sums-zeta}
\begin{split}
S_1(\mu) & = \mmfrac 12  + \floor{\mu} - \mu\,, \\
S_2(\mu) &=  - \mmfrac 1{12}- \mmfrac 12   \floor{\m} \left(\floor{\mu} + 1\right)  + \mmfrac 12 \, \mu^2\,, \\
S_3(\mu) & = \mmfrac 16 \floor{\mu} (\floor{\mu}+1)(2\floor{\mu}+1) - \mfrac 1 3 \, \mu^3\,.
\end{split}
\end{align}

Another method of regularizing the sums~\eqref{sums-ap} is to make use of the Poisson summation formula, $\sum_{n \in \Z} f(n) = \sum_{k \in \Z} \hat f(k)$, where $\hat f$ is the Fourrier transform of $f$, and then to remove the zero mode $k=0$, which captures all the divergence of the series (at least in these cases). Using the Fourrier transform of the sign function, $\hat \sgn(k) = 1/i\pi k $, and the property $\hat f (k+\m) = \hat f(k) \, {e}^{2 \pi i k \m }$, we directly obtain
\begin{align}\label{sums-pois}
\begin{split}
S_1(\mu) & = \mfrac{1}{\pi} \im \Li{1}{e^{2\pi i\m }} \\
S_2(\mu)& = - \mfrac{ \mu }{\pi} \im \Li{1}{e^{2\pi i \mu}} - \mfrac{1}{2\pi^2} \re \Li{2}{e^{2\pi i \mu}}\\
S_3(\mu)&= \mfrac{\mu^2}{\pi} \im \Li{1}{e^{2\pi i \mu}} + \mfrac{\mu}{\pi^2} \re \Li{2}{e^{2\pi i \mu}} - \mfrac{1}{2\pi^3} \im \Li{3}{e^{2\pi i \mu}}
\end{split}
\end{align}
where the Polylogarithm function $\Li{s}{z}$ is defined as $\Li{s}{z} = \sum_{k=1}^\infty z^k/k^{s}$. These expressions are valid for generic $\mu \notin \Z$ and the transformations properties~\eqref{transf-sums} follow straightforwardly from them. Using trigonometric relations, or equivalently special cases of the Hurwitz formula, it is possible to show that the results~\eqref{sums-pois} agree with those in~\eqref{sums-zeta}.

%%%%%%%%%%%%%%%%%%%%%%%%%%%%%%%%%%%%%%%%%%%%%%%%
\section{Evaluation of the Feynman diagrams} \label{ap:Feynm}
%%%%%%%%%%%%%%%%%%%%%%%%%%%%%%%%%%%%%%%%%%%%%%%%

In this appendix, we evaluate the one-loop Feynman diagrams displayed in figures~\ref{fig:a} and~\ref{fig:oneloopCSG}, with the whole KK tower of fermions running in the loops.
They contribute to the Wilsonian effective action by (in momentum space)
\begin{subequations}\label{FD}
\begin{alignat}{3}
&&\Gamma^{ab} \, A_a A_b, \qquad &&\text{with} \qquad \Gamma^{ab} &=   \sumn \intdk \mmfrac 12  \, \tr{  V^a S_k V^b S_{k+p} } \label{AAdiag}\\
%&&\Gamma^{ab}_\triangle \, \chi A_a A_b, \qquad &&\text{with} \qquad \Gamma^{ab}_\triangle &=   \sumn \intdk \mmfrac 12  \, \tr{  V^a S_{k+p} V^b S_k V S_{k-q} } \\
&&\Gamma^{ab}_0 \, A_a A^0_b, \qquad &&\text{with} \qquad \Gamma^{ab}_0 &=   \sumn \intdk \mmfrac 12  \, \tr{  V^a S_k V_0^b(-p) S_{k+p} } \\
&&\Gamma^{ab}_{00} \, A^0_a A^0_b, \qquad &&\text{with} \qquad \Gamma^{ab}_{00} &=   \sumn \intdk \mmfrac 12  \, \tr{  V_0^a(p) S_k V_0^b(-p) S_{k+p} } 
\end{alignat}
\end{subequations}
where $S_k$ is the propagator of a fermion with momentum $k$, the external fields have momentum $p$,
and $V^a$ and $V^a_0$ are the vertices between two fermions and $A_a$ and $A^0_a$ respectively. Note that $V_0$ depends on $p$, due to the Pauli term.

As they stand, the expressions~\eqref{FD} are divergent, so we need to regulate them.
We will use Pauli-Villars regularization, 
which consists in adding for each fermion extra massive ones, with possibly different statistic (i.e. without the usual minus sign for a fermionic loop), evaluate the diagrams with those extra fermion running in the loop, and finally, take their masses to infinity. In our case, three extra particles are needed, and their masses $M_s, \ s=1,2,3$, must satisfy  $\sum_{s=1}^3 (-1)^s M_s^2 =0 $. The choice $M_1^2 = \frac 12 M_2^2 = M_3^2 \equiv  M^2$ satisfies this condition, and we will make that choice in the following; we also define $M_0=0$ for ease of notation. The regularized expressions of the $\Gamma$'s in~\eqref{FD} are therefore 
\begin{subequations}\label{FDreg}
\begin{align}
 \Gamma^{ab}_\reg &=   \limms \sumn \intdk \sums \mmfrac 12  \, \tr{  V^a S^s_k V^b S^s_{k+p} }  \label{AAreg}\\
% \Gamma^{ab}_{\triangle, \reg} &=  \limms \sumn \intdk \sums \mmfrac 12  \, \tr{  V^a S^s_{k+p} V^b S^s_k V S^s_{k-q} } \label{AAchireg} \\
\Gamma^{ab}_{0,\reg} &=  \limms \sumn \intdk \sums\mmfrac 12  \, \tr{  V^a S^s_k V_0^b(-p) S^s_{k+p} } \label{AA0reg}\\
\quad \Gamma^{ab}_{00, \reg} &=  \limms \sumn \intdk \sums \mmfrac 12  \, \tr{  V_0^a(p) S^s_k V_0^b(-p) S^s_{k+p} } \label{A0A0reg}
\end{align}
\end{subequations}
where $S_k^s$ is the propagator of a PV fermion with momentum $k$ and mass $M_s$.

We will use this PV regularization through two different approaches. The first one is a purely three-dimensional regularization (which reproduces previously known results), whilst in the second one, the regulator preserves four-dimensional Lorentz invariance (leading to new results).
For the three-dimensional regularization, we must add the 3 PV massive particles to each KK-fermion, whereas in the second case, we only add them to the single four-dimensional fermion and reduce on a circle; this of course also produces a fermionic KK-tower where each mode comes with its regulators. However, as we will shortly see, the two approaches are not equivalent.
As explained in the main text, the latter is the appropriate thing to do if the theory comes from the circle reduction of a four-dimensional theory, for regularization is a process taking place in the UV of the theory.

In both approaches, we will use a ``four-dimensional formalism", that is, where the $\g$-matrices are $4\times 4$, making the comparison between the two approaches easier. 
Also, we will do everything including gravity (thus matching with section~\ref{sec:anom_circle_grav}), but if one is just interested in the coupling $A\wedge F$ (i.e. matching with section~\eqref{sec:aom_circle}), one can simply take $A^0=0$.

\subsection{3D regularization}
\label{ap:3dreg}

As just explained, for our first regularization method, we start from the action~\eqref{3dactG}, to which we add 3 PV particles for each mode, thus obtaining the following action
\begin{equation}\label{3DPV}
\begin{split}
S_3=\sum_{n \in \Z}  \int \sums \!  \, \bar \psi_{n,s} \Big \{ \gamma^a \!\left( i \partial_a + \mmfrac i2 \,\, \omega_a   \!+ q g A_a + n A^0_a - \mfrac r{8} \epsilon_{abc} F^{0bc} \right)\\
 + \mtfrac 1r \g^\bt q \chi + \gamma^\bt m_n + M_s  \Big \}P_L \psi_{n,s} \star 1 \,,
\end{split}
\end{equation}
where $m_n = \frac 1r (n +q \vz)$ in the CB.
The Feynman rules read 
  \begin{subequations} \label{feynmrules3d}
 \begin{alignat}{2}
S_k&:& \qquad\vcenter{\hbox{\rule[28pt]{5pt}{0pt}\includegraphics{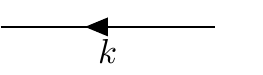}}}\!\!
 & =  P_L \,\frac{i }{\g^a k_a - \g^\bt m_n -M_s+ i\epsilon} \\
V^a&:& \qquad\vcenter{\hbox{\includegraphics[scale=1.05]{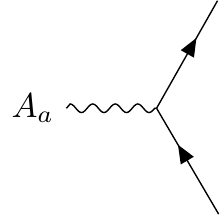}}}
\quad &= i q g \g^a P_L \\
V^a_0(p)&:& \qquad\vcenter{\hbox{\includegraphics[scale=1.05]{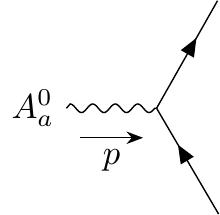}}}
\quad &= \left( i n\g^a -\mfrac r2 \,\epsilon^{abc} \g_b p_c\right) P_L %\\
%V&:&\qquad\vcenter{\hbox{\includegraphics[scale=1.05]{chiff}}}\quad &= \mtfrac 1r iq \g^\bt P_L
\end{alignat}
\end{subequations}
where $k$ is the momentum of $\psi_n$ and $p$ the one of $A^0$.
We now proceed to compute the diagrams, using these propagator and vertices for evaluating the traces in~\eqref{FDreg}.

\subsubsection[The $A-A$ diagram]{\boldmath The $A-A$ diagram}
\label{ap:AA}
Evaluating the trace in~\eqref{AAreg}, using the Feynman rules~\eqref{feynmrules3d}, and restricting to the $\e$~piece, we find
\begin{equation} \label{AAint}
\Gamma_\reg^{ab}|_\e =-i \frac{g^2}{4 \pi} \epsilon^{abc} p_c \, q^2 \limms \sumn\intk \, \sums \frac{m_n}{(k^2 - m_n^2-M_s^2+i \epsilon)^2}\,,
\end{equation}
and, performing the integral (using the choice of $M_s$ specified earlier),
\begin{equation} \label{sumPV}
\Gamma_\reg^{ab}|_\e =\frac{g^2}{4 \pi}  \epsilon^{abc} p_c \, q^2\limm \nfrac 12 \sum_{n\in \Z } \left( \sign(m_n) - \frac{2m_n}{\sqrt{m_n^2+M^2}}+\frac{m_n}{\sqrt{m_n^2+2M^2}} \right) \,.
\end{equation}
Here we pause for a moment, and look at what we would have obtained for a single KK-mode, i.e. if we drop the sum over $n$. The last two terms die in the limit $M\to \infty$ and we are left with $\frac 12 \,q^2 \sign(m_n)$, which is the result stated in~\eqref{int_ferm}.\footnote{In fact, in that case the PV regulator is not needed, the expression~\eqref{AAdiag} is already finite for a single mode.} However when the whole sum is considered, those last two terms in~\eqref{sumPV} are needed to obtain a finite result. Moreover, one first has to sum over $n$ and then take the limit $M \to \infty$. Taking opposite order, we would find the divergent sum~\eqref{sum}, which, as mentioned, can be regularized to give~\eqref{summed} (see appendix~\ref{ap:Sums}). But, as written, the sum in~\eqref{sumPV} is finite for every non-zero $M$ and taking the limit $M\to \infty$ the result agrees with~\eqref{summed}, i.e.
\begin{equation}\label{PVresAA}
\Gamma_\reg^{ab}|_\e =\frac{g^2}{4 \pi}  \epsilon^{abc} p_c \, q^2 \left(\mmfrac 12  + \floor{\mu} - \m
\right) \,,
\end{equation}
where we used the notation $\mu = q \vz$.
In the second part~\ref{ap:AA4d} of this appendix, we compute the same diagram using a regularization that preserve the 4D symmetries, and find a different result.
The same is true for the other diagrams.

\subsubsection[The $A-A^0$ diagram]{\boldmath The $A-A^0$ diagram}
\label{ap:AA0}

Evaluating the trace in~\eqref{AA0reg}, using the Feynman rules~\eqref{feynmrules3d}, and restricting to the $\e$~piece, we find
\begin{equation} \label{AA0int}
\Gamma_0^{ab}|_{\e} =-i \frac{g}{4 \pi}  \epsilon^{abc} p_c \, q\limms \sumn\intk \, \sums \frac{n\,m_n -\frac 12( m_n^2 - \nfrac 13  k^2) }{(k^2 - m_n^2-M_s^2+i \epsilon)^2}\,,
\end{equation}
where the term $n\, m_n$ comes from the minimal coupling  $\sim n A^0 \bar \psi_n \psi_n$, while $m_n^2 - \nfrac 13  k^2$ comes from the Pauli coupling $\sim F^0  \bar \psi_n \psi_n$. Using the result of the previous subsection, we see that the first one yields $\frac 12 \, q\, n \sign (m_n)$, which was expected (from eq.~\eqref{int_ferm} with one $q$ replaced by $n$). One could think that the second coupling would modify that result, but we will see that this is not the case.
Performing the integral, we find
\begin{equation} \label{sumPV0}
\begin{split}
\Gamma_0^{ab}|_\e =\frac{g}{4 \pi}  \epsilon^{abc} p_c \, q\limm \nfrac 12 \sum_{n\in \Z } \Bigg \{ n m_n \Bigg( \frac{1}{\abs{m_n}} - \frac{2}{\sqrt{m_n^2+M^2}} + \frac{1}{\sqrt{m_n^2+2M^2}}\Bigg) \\
-\frac 12 M^2 \Bigg( \frac{1}{\sqrt{m_n^2+M^2}} - \frac{1}{\sqrt{m_n^2+2M^2}}\Bigg) \Bigg \} \,.
\end{split}
\end{equation}
The first line in~\eqref{sumPV0} comes from the minimal coupling and is just obtained by replacing one of the $q$'s in~\eqref{sumPV} by an $n$. The second line comes from the Pauli coupling, and it has the sole effect of making the whole sum finite, but it is nevertheless crucial that it is taken into account. 
Performing the sum and then the limit $M \to \infty$, we find the same result as the one obtained by regularizing the second sum in~\eqref{sums}, using e.g. zeta-function regularization (see appendix~\ref{ap:Sums}), that is, 
\begin{equation} \label{}
\Gamma_0^{ab}|_\e =\frac{g}{4 \pi}  \epsilon^{abc} p_c \, q \left( - \mfrac1{12} - \mmfrac 12  \fm (\fm +1 ) + \mmfrac 12  \mu^2  \right)\,.
\end{equation}
In fact, one can already see in~\eqref{AA0int} that the new coupling does not contribute to the finite piece, using dimension regularization.\footnote{Here one cannot use PV regularization with 3 extra particles just for that piece, because the $k^2$ in the numerator makes it more divergent, and one would need to introduce more PV particles. However, we know that the whole expression has to be finite using just 3 PV particles, because the four-dimensional expression is.}

\subsubsection[The $A^0-A^0$ diagram]{\boldmath The $A^0-A^0$ diagram}
\label{ap:A0A0}

Evaluating the trace in~\eqref{A0A0reg}, using the Feynman rules~\eqref{feynmrules3d}, and restricting to the $\e$~piece, we find
\begin{equation} \label{A0A0int}
\Gamma_{00}^{ab}|_{\e} =-i \frac{g}{4 \pi}  \epsilon^{abc} p_c \limms \sumn\intk \, \sums \frac{n^2 \,m_n - n ( m_n^2 - \nfrac 13  k^2) }{(k^2 - m_n^2-M_s^2+i \epsilon)^2}\,,
\end{equation}
where, as before, the term $n\, m_n$ comes from the minimal coupling, while $m_n^2 - \nfrac 13  k^2$ comes from the Pauli coupling. It is now familiar that the first one yields $\frac 12 \, n^2 \sign (m_n)$, and that the second does not modify the result, but is nevertheless crucial to make the integral finite. Using the results of the previous section, one readily finds 
\begin{equation} \label{sumPV00}
\begin{split}
\Gamma_{00}^{ab}|_\e =\frac{g}{4 \pi}  \epsilon^{abc} p_c \limm \nfrac 12  \sum_{n\in \Z } \Bigg \{ n^2 m_n \Bigg( \frac{1}{\abs{m_n}} - \frac{2}{\sqrt{m_n^2+M^2}} + \frac{1}{\sqrt{m_n^2+2M^2}}\Bigg) \\
-n M^2 \Bigg( \frac{1}{\sqrt{m_n^2+M^2}} - \frac{1}{\sqrt{m_n^2+2M^2}}\Bigg) \Bigg \} \,.
\end{split}
\end{equation}
where again, the last line is due to the additional coupling and renders the sum finite. It is by now no surprise that performing this sum and then taking the limit $M \to \infty$, we find the same result as we obtained by regularizing the third sum in~\eqref{sums}, using e.g. zeta-function regularization (see appendix~\ref{ap:Sums}), that is, 
\begin{equation} \label{}
\Gamma_{00}^{ab}|_\e =\frac{g}{4 \pi}  \epsilon^{abc} p_c  \left( \mmfrac 16 \fm (\fm +1 ) (2\fm +1 ) - \mmfrac 13 \mu^3  \right)\,.
\end{equation}
\vspace{0mm}

\subsection{3D regularization, preserving 4D Lorentz invariance}
\label{ap:4dreg}
 
 The second method of regularization that we will use, is to add the PV particle already in four dimensions, as is done in~\eqref{S4PVG}
and then reduce on a circle. On then obtains the action~\eqref{3DPVG}, which we recall here for convenience, 
 \begin{equation}\label{3DPVapp}
\begin{split}
 S_3=\sum_{n \in \Z}  \int  \sums  \,\bar\psi_{n,s} \Big \{ \gamma^a \!\left( i \partial_a + \mmfrac i2 \,\, \omega_a P_L  \!+ q g A_a P_L + n A^0_a - \mfrac r{8 } \epsilon_{abc} F^{0bc} P_L \right) \\
  + \mtfrac 1r \g^\bt q \chi P_L + \mtfrac 1r \gamma^\bt (n + \m P_L) + M_s  \Big \} \psi_{n,s} \star 1 \,,
\end{split}
 \end{equation}
where we use the notation $\mu = \qz$.
The Feynman rules read
  \begin{subequations}\label{feynmrules4d}
 \begin{alignat}{2}
 S_k&:& \qquad\vcenter{\hbox{\rule[28pt]{5pt}{0pt}\includegraphics[scale=1.05]{prop}}}\!\! & = \frac{i }{\g^a k_a - \frac 1r \g^\bt (n + \mu P_L) -M_s+ i\epsilon} \label{prop4dPV} \\
V^a&:& \qquad\vcenter{\hbox{\includegraphics[scale=1.05]{Aff}}}
\quad &= i q g \g^a P_L \\
V^a_0(p)&:& \qquad
\vcenter{\hbox{\includegraphics[scale=1.05]{A0ff}}}
\quad &=  i n\g^a -\mfrac r2 \, \epsilon^{abc} \g_b p_c P_L \label{V0}
% \\ V&:& \qquad\vcenter{\hbox{\includegraphics[scale=1.05]{chiff}}}\quad &= \mtfrac 1r iq \g^\bt P_L
\end{alignat}
 \end{subequations}
 where $k$ is the momentum of $\psi_n$ and $p$ the one of $A^0$.
 We now proceed to compute the diagrams, using these propagator and vertices for evaluating the traces in~\eqref{FDreg}. In order to do so, one needs to invert the matrix in~\eqref{prop4dPV}. Since this propagator is non-standard, we present here the result. Defining
  \begin{align}
\begin{split}
  a_k& = k^2 - n^2 - M_s^2\\
  b_k &= k^2 - m_n^2 -M_s^2 \\
  D_k  &= a_k b_k - \m^2 M_s^2 \,.
\end{split}
  \end{align}
  (notice that $b_k$ is nothing but the denominator that appeared in the previous section, namely in eqs.~\eqref{AAint},~\eqref{AA0int}, and~\eqref{A0A0int}), we find 
 \begin{equation}
 S_k  =  \frac1{D_k}\left(\g^a  k_a + M - \g^3 (n + \m P_R) \right)\left( a_k - (\mu ^2+ 2 n \mu ) P_L + \m M \g^3 \g^5 \right) \,.
 \end{equation}
In order to compute the traces, the following identities are useful
 \begin{align}
\begin{split}
 P_L S_k P_R &= \frac{1}{D_k} P_L \slashed \xi _k, 
 \quad \text{with} \quad\xi_{k,\bm} =(a_kk_\ba,-a_k m_n-\m M^2)  \,,\\
 P_R S_k P_L &= \frac{1}{D_k}  \slashed \zeta_k P_L, 
 \quad \text{with} \quad\zeta_{k,\bm} =(b_kk_\ba,-b_kn+\mcb M^2) \,, \\
 P_L S_k P_L &= \frac{M}{D_k} P_L \left( \mcb  \g^a \g^3 k_a +  c_k \right), \quad \text{with} \quad c_k = a_k - n \mcb \,,\\
 P_R S_k P_R &= \frac{M}{D_k} P_R \left( \mcb \g^3 \g^a k_a +  c_k\right) \,,
\end{split}
 \end{align}
 where the slashes mean contraction with four-dimensional flat gamma matrices $\slashed k = \g^\bm k_\bm$.

 % % % % % % % % % % % % % % % % % % % % % % % % % % % % % % %
\subsubsection[The $A-A$ diagram]{\boldmath The $A-A$ diagram}
\label{ap:AA4d}

Evaluating the trace in~\eqref{AAreg}, using the Feynman rules~\eqref{feynmrules4d}, and restricting to the $\e$~piece, we find\footnote{The bold subscript \textbf{reg} is to indicate that we use a different regularization, as mentioned in the main text.}
\begin{equation}\label{AA4D}
\Gamma_\regb^{ab}|_\e = -i \ffpi \epsilon^{abc} p_c \, q^2  \limms \sumn \intk \sums \frac{1}{D_k^2} \left[a_k^2 \, m_n +  \m M_s^2 \left (a_k + \mmfrac 23 k^2 \right )\right]\,.
\end{equation}
This result agrees with~\eqref{AAint} for the term $s=0$, as it should, since this term corresponds to the physical particles, and therefore should not depend on the regularization. However it differs from~\eqref{AAint} for $s \neq 0$, meaning that the two different ways of regularizing do not yield the same results. But notice that by setting $\mu=0$ everywhere in~\eqref{AA4D}, except in $m_n$, we recover precisely~\eqref{AAint}.\footnote{To see this, notice that $a_k/D_k =1/b_k + \c O (\m^2)$.} This means that by expanding~\eqref{AA4D} in powers of $\mu$, one finds~\eqref{AAint} plus a series in $\mu$ starting at order one, 
\begin{equation}\label{AA4D2}
\Gamma_\regb^{ab}|_\e = -i \ffpi \epsilon^{abc} p_c \, q^2  \limms \sumn \intk \sums \left( \frac{m_n}{b_k^2} +   \sum_{p=1}^\infty \varGamma_p \right)\,.
\end{equation}
where $\varGamma_p$ is order $p$ in $\mu$. The first term was already computed in section~\ref{ap:AA} and gives~\eqref{PVresAA}, namely we have
\begin{equation}\label{Acontr}
-i \limms \sumn \intk \sums 
 \frac{m_n}{b_k^2}  = \left(\mmfrac 12  + \floor{\mu} - \m
\right) \,.
\end{equation}
Concerning the second part, only the first term of the series survives in the limit $M_s\to \infty$. It reads
\begin{equation}
\varGamma_1 =  \frac{\mu M_s^2}{ a_k^4} \left (\mfrac 53 \,k^2 -n^2 -M_s^2 \right ) \,.
\end{equation}
Performing the integral over $k$, we find 
\begin{equation}
-i \intk \sums \varGamma_1= \frac{1}{3} \mu M^2 \left ( \frac 1{(n^2+M^2)^{3/2}} - \frac 1{(n^2+ 2M^2)^{3/2}} \right ) \,.
\end{equation}
In the limit $M \to \infty$, the sum over $n$ can be converted into an integral by the rescaling $n = M L$, and we find 
\begin{equation}\label{Bcontr}
\limms \intk \sums \varGamma_1= \nfrac 13  \mu  \int_{-\infty}^\infty \d L \left( \frac{1}{(L^2 + 1)^{3/2}} - \frac{1}{(L^2 + 2)^{3/2}}\right) = \nfrac 13  \mu\,.
\end{equation}
Combining~\eqref{Acontr} with~\eqref{Bcontr}, we find
\begin{equation}
\Gamma_\regb^{ab}|_\e = \ffpi \epsilon^{abc} p_c \, q^2 \left(\mmfrac 12  + \floor{\mu} - \mmfrac 23\,\m \!\right) \,,
\end{equation}
which is the result stated in~\eqref{PVregG}. Note that the difference between this regularization and the one of section~\ref{ap:AA}, i.e.~\eqref{Bcontr}, is precisely the anomaly inflow term~\eqref{AI}, as was expected.

% % % % % % % % % % % % % % % % % % % % % % % % % % % % %
\subsubsection[The $A-A^0$ diagram]{\boldmath The $A-A^0$ diagram}
\label{ap:AA04d}

Evaluating the trace in~\eqref{AA0reg}, using the Feynman rules~\eqref{feynmrules4d}, and restricting to the $\e$~piece, we find
\begin{equation}
\Gamma^{ab}_{0,\regb}|_\e  = -i \frac{g}{4\pi} \epsilon^{abc} p_c q \limms \sumintk \eta_s \left (   \varGamma_{0,n}  + \varGamma_{0,\e}\right )
\end{equation}
where
\begin{align}
\Gamma_{0,n} &= \frac{n}{D_k^2} \left(a_k^2 m_n  + 2 a_k\mcb M_s^2 - n \mcb^2 M_s^2\right) \notag \\
\Gamma_{0,\e} & =  \frac{1}{2 D_k^2}\left(\mfrac {1}3 a_k^2 k^2  - (a_k m_n + \mcb M_s^2)^2\right) \notag
\end{align}
$\varGamma_{0,n}$ comes from the first coupling in~\eqref{V0} and $\varGamma_{0,\e}$ comes from the second one. 
Here again, the case $s=0$ coincide with~\eqref{AA0int} and expanding in $\mu$, we find the corresponding term, plus a series starting at first oder in $\mu$
\begin{equation}
\Gamma^{ab}_{0,\regb}|_\e =  -i \frac{g}{4\pi}\epsilon^{abc} p_c \, q \limms\sumintk \eta_s \bigg [ 
\frac{n\,m_n -\frac 12( m_n^2 - \nfrac 13  k^2) }{(k^2 - m_n^2-M_s^2)^2}
+   \sum_{p=1}^\infty \varGamma_{0,p} \,\bigg] \,.
\end{equation}
The first piece was computed in section~\eqref{ap:AA0} and yields
$- \mmfrac 1{12} - \mmfrac 12  \fm (\fm +1 ) + \mmfrac 12  \mu^2$,
while for the second piece, only the $p=2$ term of the series contributes, and we find
\begin{equation}
-i \limms \sumintk \eta_s    \varGamma_{0,2} = - \nfrac 13  \,  \mu^2\,,
\end{equation}
which is again the anomaly inflow piece. The total result thus is
\begin{equation}
\Gamma^{ab}_{0,\regb}|_\e =  \frac{g}{4\pi}\,\epsilon^{abc} p_c \, q \left(- \mmfrac 1{12} - \mmfrac 12  \fm (\fm +1 ) + \mmfrac 16 \, \mu^2\right)\,,
\end{equation}
which is the result stated in~\eqref{PVregG}.

% % % % % % % % % % % % % % % % % % % % % % % % % % % % % % % % % % %
\subsubsection[The $A^0-A^0$ diagram]{\boldmath The $A^0-A^0$ diagram}
\label{ap:A0A04d}
Evaluating the trace in~\eqref{A0A0reg}, using the Feynman rules~\eqref{feynmrules4d}, and restricting to the $\e$-piece, we find
\begin{equation}
\Gamma^{ab}_{00,\regb}|_\e  = -i \frac{1}{4\pi} \epsilon^{abc} p_c \limms \sumintk \eta_s \left (   \varGamma_{0,nn}  + \varGamma_{00,n\e}\right )
\end{equation}
where
\begin{align}
\Gamma_{00,nn} & = \frac{n^2}{D_k^2} \left(a_k^2 m_n  - b_k^2 n + 4 a_k\mcb M^2 - 4 n \mcb^2 M^2- \mcb^3 M^2 \right) \notag \\
\Gamma_{00,n\e} & =  \frac{n}{D_k^2} \left ( \mfrac {1}3 (a_k^2 + \mcb^2 M^2) k^2  - (a_k m_n + \mcb M^2)^2 - M^2 c_k^2 \right ) \notag
\end{align}
$\varGamma_{00,nn}$ comes from considering the minimal coupling $\sim n$ in~\eqref{V0} for both vertices and $\varGamma_{00,n\e}$ comes from having the Pauli coupling in one of them (taking twice this coupling yields a term with 2 $p$'s, which is thus higher derivative).
Once more, the case $s=0$ coincides with~\eqref{A0A0int} and expanding in $\mu$, we find the corresponding term, plus a series starting at first oder in~$\mu$
\begin{equation}
\Gamma^{ab}_{00,\regb}|_\e =  -i \frac{1}{4\pi}\epsilon^{abc} p_c \limms\sumintk \eta_s \bigg [ 
\frac{n^2 \,m_n - n ( m_n^2 - \nfrac 13  k^2) }{(k^2 - m_n^2-M_s^2)^2}
+   \sum_{p=1}^\infty \varGamma_{00,p} \,\bigg] \,.
\end{equation}
The first piece was computed in section~\eqref{ap:A0A0} and yields
$\mmfrac 16 \fm (\fm +1 ) (2\fm +1 ) - \mmfrac 13 \mu^3 $,
while for the second piece, only the $p=3$ term of the series contributes, and we find
\begin{equation}
-i \limms \sumintk \eta_s    \varGamma_{00,3} =  \nfrac 13 \,  \mu^3\,,
\end{equation}
which is again the anomaly inflow piece. The total result thus is
\begin{equation}
\Gamma^{ab}_{00,\regb}|_\e =  \frac{1}{4\pi}\,\epsilon^{abc} p_c \, \, \mmfrac 16 \, \fm (\fm +1 ) (2\fm +1 ) \,,
\end{equation}
which is the result stated in~\eqref{PVregG}.

\renewcommand{\baselinestretch}{1.1} 
\bibliographystyle{JHEP}
\bibliography{anomalies.bib}

\end{document}